\documentclass[bibyear]{aa}
\usepackage{graphicx}
\usepackage{color}
\usepackage{txfonts}\usepackage{subcaption}
\usepackage{natbib}
\usepackage{gensymb}
\usepackage{graphicx}
\usepackage{txfonts}
\usepackage{newtxtext,newtxmath}
\usepackage{ae,aecompl}
\usepackage{amsmath}	% Advanced maths commands
\usepackage{amssymb}	% Extra maths symbols
\usepackage{float}
\usepackage{pdflscape}
\everymath{\displaystyle}

\begin{document}

	\title{Tracking the spectral properties across the different epochs in ESO 511-G030}
	\titlerunning{The ESO 511-G030 spectrum across the years}
	\authorrunning{ R. Middei at al.,}
	
	\subtitle{}%A serendipitous broadband X-ray spectral analysis}
	
   \author{R. Middei
		\inst{1,2}\fnmsep\thanks{riccardo.middei@ssdc.asi.it}
		P.-O. Petrucci \inst{3}, S. Bianchi\inst{4},   F. Ursini \inst{4} G. A. Matzeu \inst{5,6,7}, F. Vagnetti\inst{8,9}, A. Tortosa \inst{10}, A. Marinucci \inst{11}, G. Matt\inst{4}, E. Piconcelli \inst{2}, A. De Rosa \inst{12}, B. De Marco \inst{13}, J. Reeves\inst{14,15}, M. Perri \inst{1,2}, M. Guainazzi \inst{16}, M. Cappi \inst{6}, C. Done \inst{17}}
		
	    \institute{Space Science Data Center, SSDC, ASI, via del Politecnico snc, 00133 Roma, Italy
		\and INAF - Osservatorio Astronomico di Roma, via Frascati 33, I-00040 Monteporzio Catone, Italy
		\and Univ. Grenoble Alpes, CNRS, IPAG, F-38000 Grenoble, France
		\and Dipartimento di Matematica e Fisica, Universit\`a degli Studi Roma Tre, via della Vasca Navale 84, I-00146 Roma, Italy
		\and Department of Physics and Astronomy (DIFA), University of Bologna, Via Gobetti, 93/2, I-40129 Bologna, Italy
		\and INAF-Osservatorio di Astrofisica e Scienza dello Spazio di Bologna, Via Gobetti, 93/3, I-40129 Bologna, Italy
		\and European Space Agency (ESA), European Space Astronomy Centre (ESAC), E-28691 Villanueva de la Cañada, Madrid, Spain
		\and INAF - Istituto di Astrofisica e Planetologia Spaziali, Via del Fosso del Cavaliere, 00133, Roma, Italy
		\and Dipartimento di Fisica, Universit\`a degli Studi di Roma Tor Vergata, Via della Ricerca Scientifica 1, 00133,Roma, Italy 
		\and N\'ucleo de astronom\'ia de la facultad de ingenier\'ia, Universidad Diego Portales, Av. Ej\'ercito Libertador 441, Santiago, Chile
		\and ASI - Italian Space Agency, Via del Politecnico snc, 00133, Rome, Italy
		\and INAF/IAPS - Istituto di Astrofisica e Planetologia Spaziali, Via del Fosso del Cavaliere I-00133, Roma, Italy
		\and Departament de F\`isica, EEBE, Universitat Politècnica de Catalunya, Av. Eduard Maristany 16, E-08019 Barcelona, Spain
		\and Department of Physics, Institute for Astrophysics and Computational Sciences, The Catholic University of America, Washington, DC 20064, USA 
    	\and INAF - Osservatorio Astronomico di Brera, Via Bianchi 46 I-23807 Merate (LC), Italy
    	\and European Space Agency, ESTEC, Keplerlaan 1 2201AZ Noordwijk, The Netherlands
    	\and Centre for Extragalactic Astronomy, Department of Physics, University of Durham, South Road, Durham DH1 3LE, UK
	}
	
	%   \date{Received September 15, 1996; accepted March 16, 1997}
	
	% \abstract{}{}{}{}{} 
	% 5 {} token are mandatory
	
	\abstract
	% context heading (optional)
	% {} leave it empty if necessary  
	{The Type I active galactic nucleus (AGN) ESO 511-G030, a formerly bright and soft-excess dominated source, has been observed in 2019 in the context of a multi-wavelength monitoring campaign. However, in these novel exposures, the source was found in a $\sim$10 times lower flux state, without any trace of the soft-excess. Interestingly, the X-ray weakening corresponds to a comparable fading of the UV suggesting a  strong link between these components. The UV/X-ray spectral energy distribution (SED) of ESO 511-G030 shows remarkable variability. We tested both phenomenological and physically motivated models on the data finding that the overall emission spectrum of ESO 511-G030 in this extremely low flux state is the superposition of a power-law-like continuum ($\Gamma\sim$1.7) and two reflection components emerging from hot and cold matter. has Both the primary continuum and relativistic reflection are produced in the inner regions. The prominent variability of ESO 511-G030 and the lack of a soft-excess can be explained by the dramatic change in the observed accretion rate, which dropped from an L/L$_{\rm Edd}$ of 2\% in 2007 to 0.2\% in 2019. The  X-ray photon index also became harder during the low flux 2019 observations, perhaps as a result of a photon starved X-ray corona.}
	

	\maketitle
	%
	%-------------------------------------------------------------------
	
	\section{Introduction}
	
	The broadband emission in active galactic nuclei (AGNs) can be ascribed as an interplay between thermal and non-thermal processes taking place in the close surroundings of a central supermassive black hole \citep[SMBH; see][for an overview on AGNs]{Padovani17}. Accretion onto the SMBH is responsible for the optical-UV emission and a fraction of these thermal photons is intercepted and Comptonised up to the X-rays by the so-called hot corona \cite[e.g.][]{Galeev1979,haar91,haar93}. Our knowledge on this plasma is still lacking, though timing and microlensing arguments \citep[e.g.][]{Chartas09,Morgan12,DeMa13,Kara16} support the hypothesis of this component being compact and located in the inner regions of the accretion flow.\\
	\indent The presence of this hot plasma well explains the cut-off power-law like continuum of AGNs with the high energy roll over being interpreted as a further signature of the inverse-Compton of seed photons by the thermal plasma of electrons with E$_{\rm c}$ depending on the temperature of the relativistic electrons (kT) \citep[][]{Rybi79}. 
     The characterisation of the primary X-ray continuum has been the focus of several studies \citep[e.g.][]{Pero00,Dadina07,Molina09,Molina13,Mali14,Ricci18} and it has been boosted after the launch of \textit{NuSTAR} \cite[][]{Harrison2013} as demonstrated by the increasing number of high energy cut-off measurements \cite[e.g.][]{Balokovic2020,Reeves2021,Kamraj2022}.\\
	\indent Reflection off Compton-thin/thick matter imprints additional features to the emerging X-ray spectrum. This is the case of the fluorescence Fe K$\alpha$ emission line at 6.4 keV \citep[e.g.][]{Bian09} or the Compton-hump \citep[][]{George91,Matt93,Garc14}. Noticeably, the analysis of the Fe K$\alpha$ profile carries a wealth of information on the location of the reflecting materials. In fact, its intrinsically narrow profile, can undergo distortions, resulting into a broader shape, due to relativistic effects \citep[e.g.][]{Fabian1989,Tanaka1995,Fabian2000,Nandra2007,deLaCalle2010}. \\
	\indent Finally the soft X-ray band of AGNs ubiquitously show a bump of counts below $\sim$2 keV that is not accounted by the high energy power-law continuum \cite[e.g][]{Piconcelli2005,Bian09,Gliozzi2020}. The origin of this component, the so-called soft-excess, is still debated and two possible scenarios have been tested on various data-sets: blurred ionised reflection and warm Comptonisation \citep[e.g.][]{Crummy06,Magdziarz1998,Jin2012,Done2012}. Both these models have been found to  reproduce the data: \cite{Walton2013} tested the relativistic reflection origin of the soft-excess on a broad number of Seyfert galaxies using \textit{Suzaku} data \citep[see also][]{Crummy06}, while the so-called two-coronae model was found to best-fit the broadband emission spectrum of an increasing number of AGNs \citep[e.g.][]{Petrucci2018,Porquet2018,Kubota2018,Ursini18,Ursini2020,Mahmoud2020,Matzeu2020,Middei2020}.\\
	\indent Variability is another key feature of AGNs emission \citep[][]{Bregman1990,Mushotzky1993,Wagner1995,Ulrich1997}. Spectral variations are characterised by a softer-when-brighter behaviour that is commonly observed in nearby Seyfert galaxies and unobscured quasars \citep[e.g.][ respectively]{Sobolewska2009,Serafinelli2017}. X-ray amplitude variations have been witnessed over different time-intervals, from hourly \citep{Ponti12} up to yearly changes \citep[][]{Vagnetti2016,Paolillo2017,Middei2017,Falocco2017,Gallo2018,Timlin2020}.  X-ray variability has been found to anti-correlate with the source luminosity, with this being naturally explained if changes result by the superposition of $N$ randomly emitting sub-units. This scenario, already considered in optical studies \citep[e.g.][]{Pica1983,Aretxaga1997}, predicts a variability amplitude $\propto N^{-0.5}\propto L^{-0.5}$ and accounts for the sub-units to be identical and flare independently \citep[e.g.][]{Green1993,Almaini2000}. Interestingly, this behaviour has been reported by many authors, both for local and high-redshift AGNs also in the X-rays \citep[e.g.][]{Barr1986,Lawrence1993,Papadakis2008,Vagnetti2011,Vagnetti2016}.\\
	\indent In this context, we report on the X-ray spectral properties of ESO 511-G030, a nearby \citep[$z=0.02239\pm0.00001$,][]{Theureau1998} spiral galaxy \citep[][]{deVaucouleurs1991} hosting an unobscured type 1 AGN \citep[][]{Veron2010}. 
	One of the first X-ray spectra of ESO 511-G030 was taken with \textit{ASCA} in 1998 \citep[]{Turner2001asca} in which the source spectrum was best-fitted using an absorbed power-law, while a subsequent \textit{INTEGRAL}-based analyses allowed to measure a high energy cut-off E$_{\rm c}$=100$^{+101}_{-37}$ keV \citep{Mali14}. The AGN ESO 511-G030 is reported in the 105-month BAT catalog \citep[][]{Oh2018}, with a flux of $\sim$4$\times$10$^{-11}$ erg s$^{-1}$ cm$^{-2}$ (14-195 keV) and is one of the 13 objects in the \textit{FERO} sample, an \textit{XMM-Newton}-based collection of AGNs with a >5$\sigma$ detection of a relativistic iron line \citep[details in][]{deLaCalle2010}.  In a recent paper by \citet[][]{Ghosh2020}, who studied a 2007 \textit{XMM-Newton} exposure, the source flux was consistent with F$_{2-10~keV}\sim$2$\times$ 10$^{-11}$ erg s$^{-1}$ cm$^{-2}$. In the same paper, two \textit{Suzaku} observations, five years apart from the \textit{XMM-Newton} one, are also discussed. In those observations, the source flux was compatible with the \textit{XMM-Newton} 2007 observation and, in the second one, about 50\% higher. Moreover, ESO 511-G030 do not show the presence of cold and/or warm absorption components \citep[e.g.][]{Laha2014}, leaving the soft X-ray band free from complex absorption features, and only a modest attenuation of the UV emission.\\
	\indent The paper is organised as follows: Sect. 2 reports on data reduction and in Sect. 3 the timing properties of the observations are discussed.  In Sect. 4 we describe the spectral analyses of the \textit{XMM-Newton/NuSTAR} 2019 monitoring campaign where we test the same spectral model on the archival data in Sect. 5. Then Sect. 6 and Sect. 7 describe the broadband {\it Swift} data analysis including {\it XRT} and {\it UVOT} data. In Sect. 8 we tested a self-consistent model on the 2019 and the 2007 {\it XMM-Newton} exposures also considering optical monitor ({\it OM}) data. Our conclusions and comments are reported in final Sect. 9.

	\section{Data reduction}
		\begin{table}
		\centering
		\caption{\small{Log of the observations. The $\dagger$ identify exposures from the joint \textit{XMM-Newton}-\textit{NuSTAR} monitoring campaign}\label{log}}
		\begin{tabular}{c c c c}
			\hline
			Observatory & Obs. ID & Start date & Net exp. \\
			& & yyyy-mm-dd & ks \\
			\hline 
			\textit{ASCA} &76067000  &1998-02-06 &17\\
			\hline
			\textit{XMM-Newton} &0502090201  &2007-08-05 & 120\\
			\hline \textit{Suzaku} &707023010  &2012-07-20& 5.7\\
			\hline \textit{Suzaku} &707023020  &2012-07-22 & 224\\
			\hline \textit{Suzaku} &707023030  &2012-08-17 & 51\\
			\hline
			\textit{XMM-Newton}$\dagger$ &0852010101  &2019-07-20 & 35\\
			\textit{NuSTAR} & 60502035002 & 2019-07-20 & 52 \\
			\hline
			\textit{XMM-Newton}$\dagger$ &0852010201  & 2019-07-25  & 37\\
			\textit{NuSTAR} & 60502035004 & 2019-07-25 & 49 \\
			\hline
			\textit{XMM-Newton$\dagger$} &0852010301  &	
			2019-07-29 & 35 \\
			\textit{NuSTAR} &  60502035006&2019-07-29& 51 \\
			\hline
			\textit{XMM-Newton}$\dagger$ &0852010401  &	
			2019-08-02  & 40 \\
			\textit{NuSTAR} &  60502035008&2019-08-02 & 48 \\
			\hline
			\textit{XMM-Newton}$\dagger$ &0852010501  & 2019-08-09 & 35 \\
			\textit{NuSTAR} &   60502035010&2019-08-09& 51 \\
			\hline
		\end{tabular}
	\end{table}

\begin{figure*}
\centering
\includegraphics[width=0.99\textwidth]{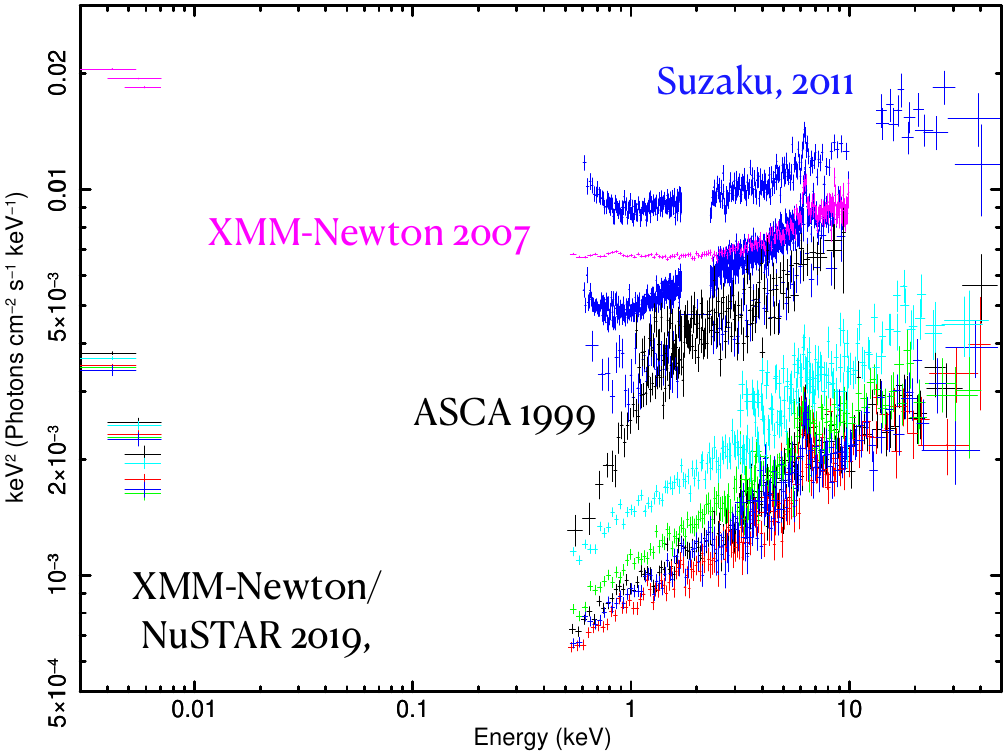}
\caption{\small{Unfolded spectra of ESO 511-G030 corresponding to the observations quoted in Table~\ref{log}. Variability is observed both in terms of spectral and amplitude changes. In the optical/UV band the source faded by a factor of $\sim$10. The underlying model folding the data is a power-law  with $\Gamma$=2  and a unitary normalisation.}\label{initialfig}}
\end{figure*}
\begin{figure*}[]
	\centering
	\includegraphics[width=0.99\textwidth]{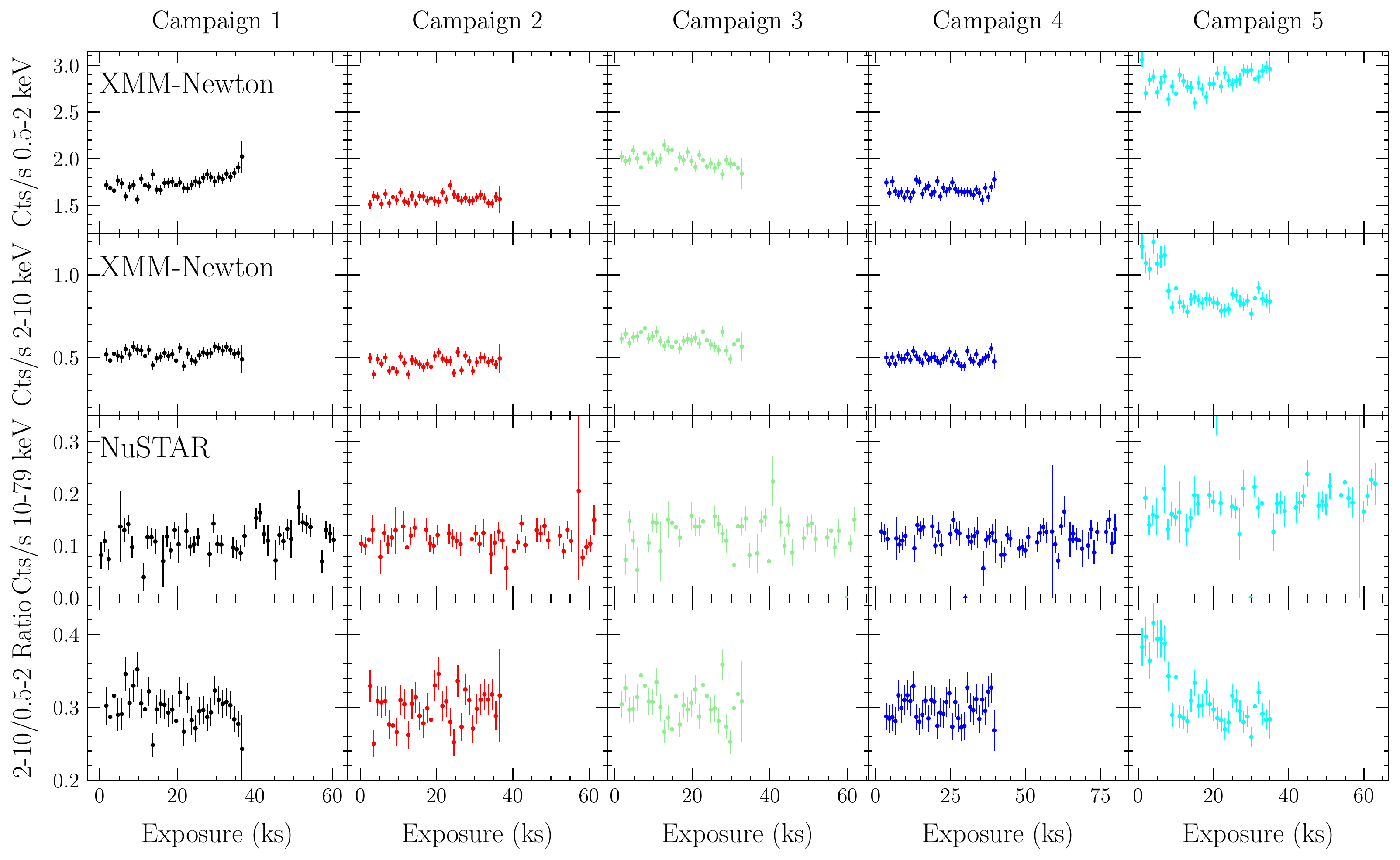}
	\includegraphics[width=0.99\textwidth]{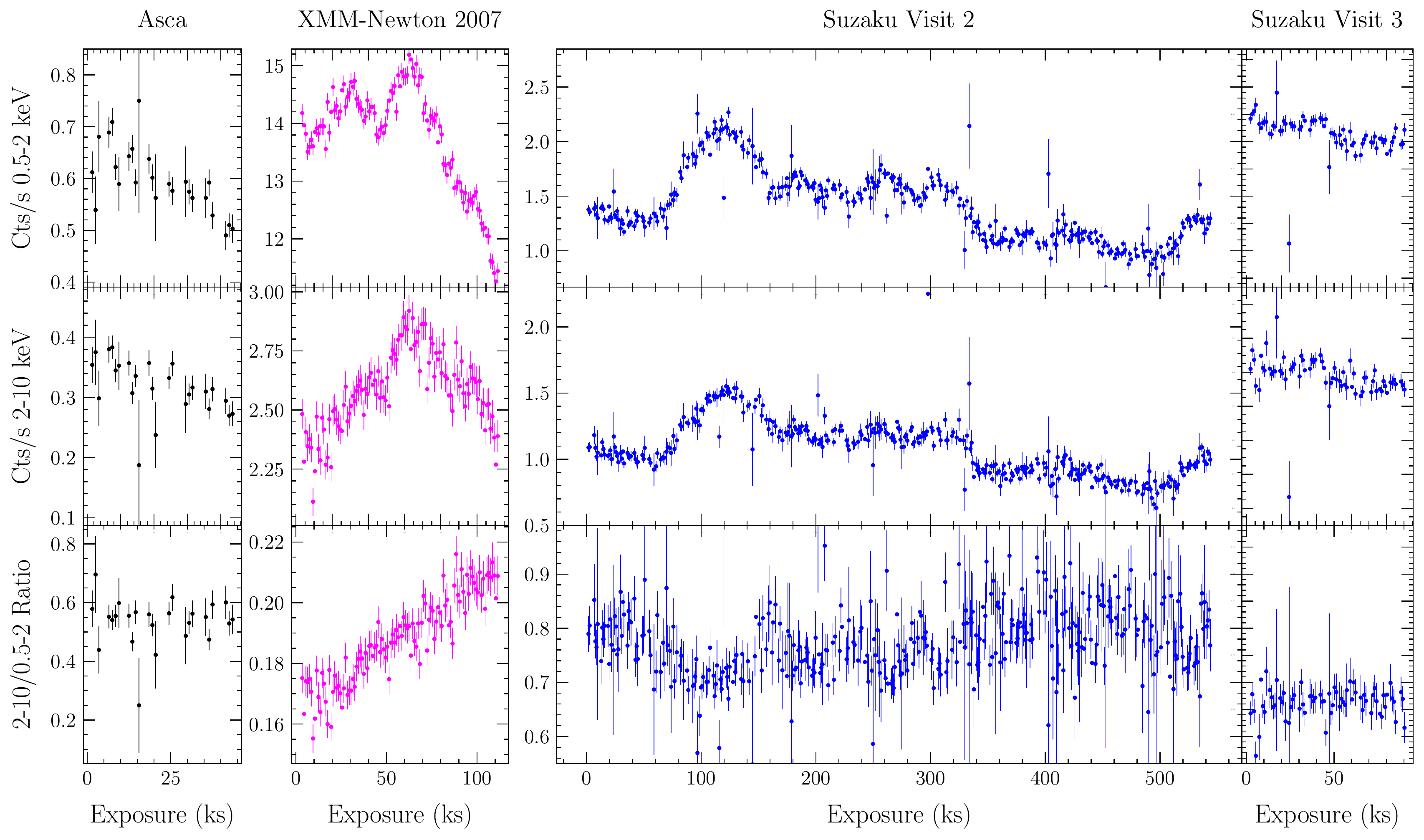}
	\caption{\small{Top panels: Background subtracted light curves from the 2019 multiwavelength campaign. The first two rows refer to the \textit{XMM-Newton} 0.5-2 and 2-10 keV, respectively, while the last one shows the ratios between these two bands. Finally, \textit{NuSTAR} light curves extracted in the 10-79 keV band are showed. Bottom panels: background subtracted time series for ASCA (black), the 2007  \textit{XMM-Newton} orbit (red) and the Suzaku visits 2 and 3 (in blue).\label{lc}}}
\end{figure*}

\begin{figure}
	\centering
	\includegraphics[width=0.45\textwidth]{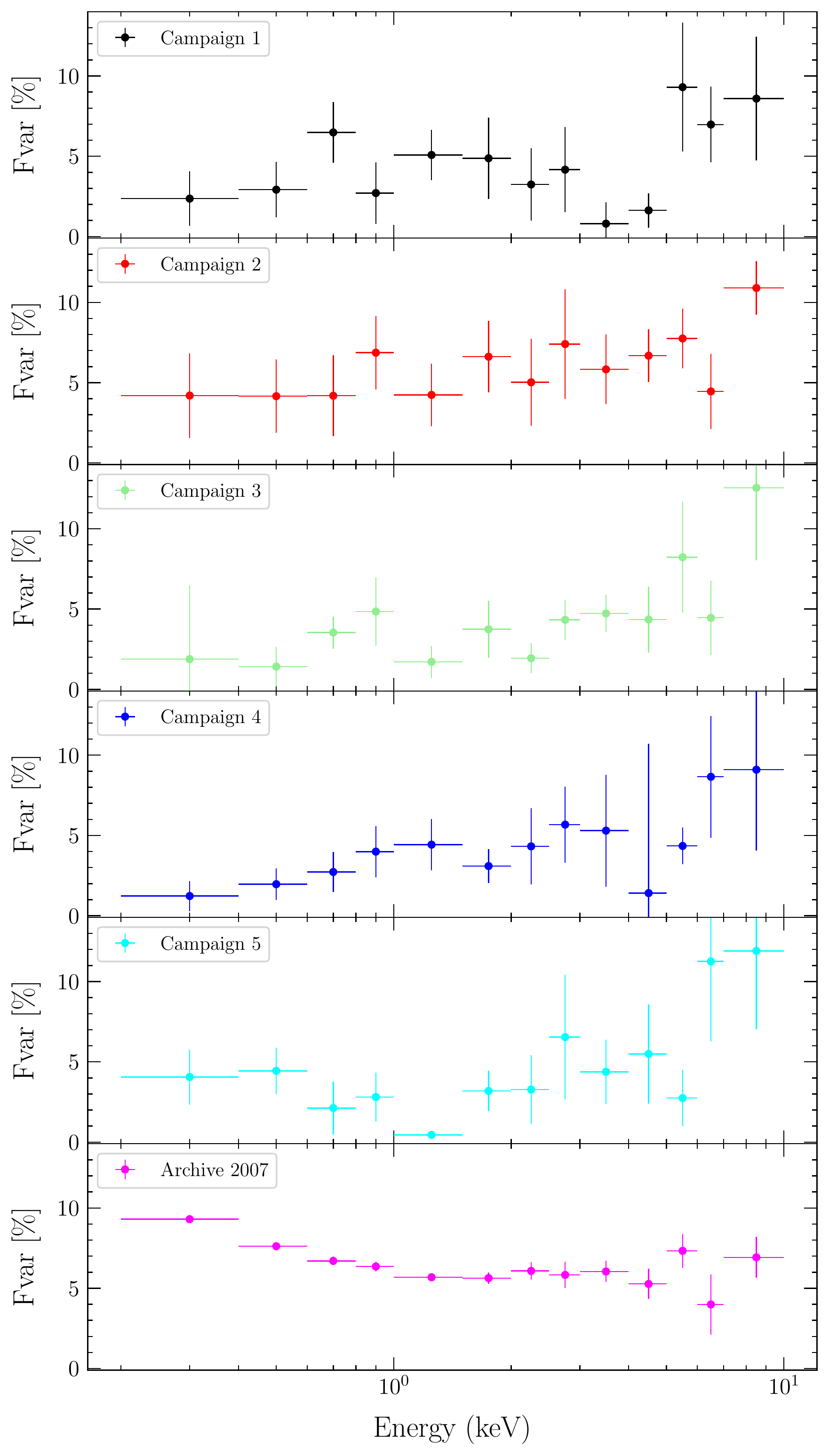}
	\caption{\small{F$_{\rm var}$ spectra for the  \textit{XMM-Newton} observations in Table~\ref{log}. Spectra from the 2019 monitoring campaign are all characterised by a fairly flat shape while data from 2007 clearly show a variability bump in the soft X-rays.\label{fvar}}}
\end{figure}
	We present the analysis of our multi-wavelength \textit{XMM-Newton/NuSTAR} observational campaign from 2019 and we compare it with data taken using different facilities across a time interval longer than 20 years. In Table~\ref{log} the log of the observations is reported while in Fig.~\ref{initialfig} we show a quick look of all spectra simply folded with a power-law ($\Gamma$=2), from which it is possible to witness the remarkable variations of ESO\,511-G030 in both the X-ray and optical/UV bands.\\
	\begin{itemize}
	\item \textit{ASCA}: We retrieved the already reduced data products of ESO 511-G030 from the Tartarus ASCA AGN database \citep[]{Turner2001asca}\\
	\item \textit{Suzaku}:
	ESO\,511-G030 was observed with \textit{Suzaku} \citep{Mitsuda2007} on July 20th (OBSID: 707023010), 22th (707023020) and August 6th (707023030) 2012 through the X-ray Imaging Spectrometer \citep[XIS;][]{Koyama2007} for net exposure times of $5.7\,\rm ks$, $224\,\rm ks$ and $51\,\rm ks$ respectively. Following the processes described in the \textit{Suzaku} data reduction guide\footnote{http://heasarc.gsfc.nasa.gov/docs/suzaku/analysis/abc}, the XIS\,0, 1 and 3 CCD spectra were extracted with \textit{HEASOFT} (v6.29.1) and adopting the latest version of the \textit{CALDB} (November 2021). 	
	The cleaned event files were selected from the 3$\times$3 and 5$\times$5 edit modes and subsequently processed according to the suggested screening criteria. Both XIS source and background spectra were extracted from circular regions of radius $3.0$\,arcmin. Care was also taken in avoiding the chip corners containing the Fe$^{55}$ calibration sources. The corresponding spectra and lightcurves were subsequently extracted using \textsc{xselect} for the second and third observations as the first pointing is too short. For each detector, the response matrices (RMF) and the ancillary response files (ARF) were generated by running the \textsc{xisrmfgen} and \textsc{xissimarfgen} tasks. After verifying their consistency, we combined the front illuminated (XIS-FI) 0 and 3 spectra into a single XIS03-FI spectrum for observations 707023020 and 707023030.
	We used a cross-calibration constant to account for the inter-calibration between the XIS and Pin detectors. In the fits, this constant has its expected value k$\sim$1.16 for the 707023030 data only while it goes to a value of k$\sim$1.50 in the 707023020 observation. Such a particularly high value for this constant is explained by the non-simultaneity of the XIS-PIN exposures due to telemetry issues which occurred leaded to an actual shortening of the PIN exposure to about 1/5 of what was scheduled.
	Then, the high value of the cross-correlation constant is straightforwardly explained by the intra-observation variability that the source had undergone.
	 \\
	\item \textit{XMM-Newton}: We reduced and analysed both a 120 ks 2007 \textit{XMM-Newton} orbit and the five exposures about 30 ksec each that  were obtained simultaneously with \textit{NuSTAR} in 2019. The exposures were performed with the \textit{EPIC} camera \citep{Struder2001,Turner2001} operating in the Small Window mode. Data were processed using the  \textit{XMM-Newton} Science Analysis System (SAS, Version 19.0.0). Because of its larger effective area with respect to the \textit{MOS} cameras, we only report the results for the \textit{pn} instrument. Source spectra were derived using a circular region with a 40 arcsec radius centered on the source while the background was extracted from a blank 50 arcsec radius area nearby the source. The extraction regions were selected using an iterative process that maximises the S/N similarly to what described in \citet{Piconcelli2004}. The spectra were rebinned
in order to have at least 30 counts for each bin and not to over-sample the spectral resolution by a factor greater than 3. Finally, from the \textit{epatplot}, pile-up issues are not affecting this dataset.
\item \textit{NuSTAR}: We calibrate and clean raw \textit{NuSTAR} \citep{Harr13}  data using the NuSTAR Data Analysis Software (\textit{NuSTARDAS}, Perri et al., 2013\footnote{\url{https://heasarc.gsfc.nasa.gov/docs/nustar/analysis/nustar\_swguide.pdf}}) package (v. 1.8.0). Level 2 cleaned products were obtained with the standard \textit{nupipeline} task while 3rd level science products were computed with the \textit{nuproducts} pipeline and using the calibration database 20191219. A circular region with a radius of 50 arcsec was used to extract the source spectrum. The background has been calculated using the same circular region but centered in a blank area nearby the source. To account for the inter-calibration of the two modules carried on the {\it NuSTAR} focal plane, we used in all the fits a cross-normalisation constant. Such a calibration constant was always found to be within 3\% this indicating the \textit{FPMA/B} spectra to be in good agreement. Spectra were binned so that each bin has at least 50 counts  and to not over-sample the
instrumental resolution by a factor greater than 2.5.\\

\item \textit{Swift}: the satellite observed  ES0511-G030  from 2018 to 2021 and we reduced data acquired with XRT and UVOT. The X-ray telescope XRT observed the source in photon counting mode and we derived  the  corresponding scientific  products  using  the  facilities  provided  by the Space Science Data Center, (SSDC,https://www.ssdc.asi.it/) of the Italian Space Agency (ASI). In particular,  spectra were extracted adopting a circular region of $\sim$60 arcsec centered on the source and a concentric annulus was used for the background. Then spectra have been binned in order to have at least 5 counts in each bin. The UVOT aperture photometry was used then to obtain the monochromatic fluxes for all the available filters. A source extraction region of 5 arcsec radius was adopted and an appropriate blank annular region concentric with the source was adopted for the background.\\
\end{itemize}
\indent  All the errors quoted in Tables and text account for 90\% uncertainties, while 68\% errors are shown in the plots. Fits are performed using \textit{Xspec} \citep[][]{Arna96} assuming the standard cosmological framework given by  H$_0$ = 70 Km s$^{-1}$  Mpc$^{-1}$, $\Omega_\Lambda$=0.73, and $\Omega$m=0.27.

\begin{figure}[t]
	\centering
	\includegraphics[width=0.45\textwidth]{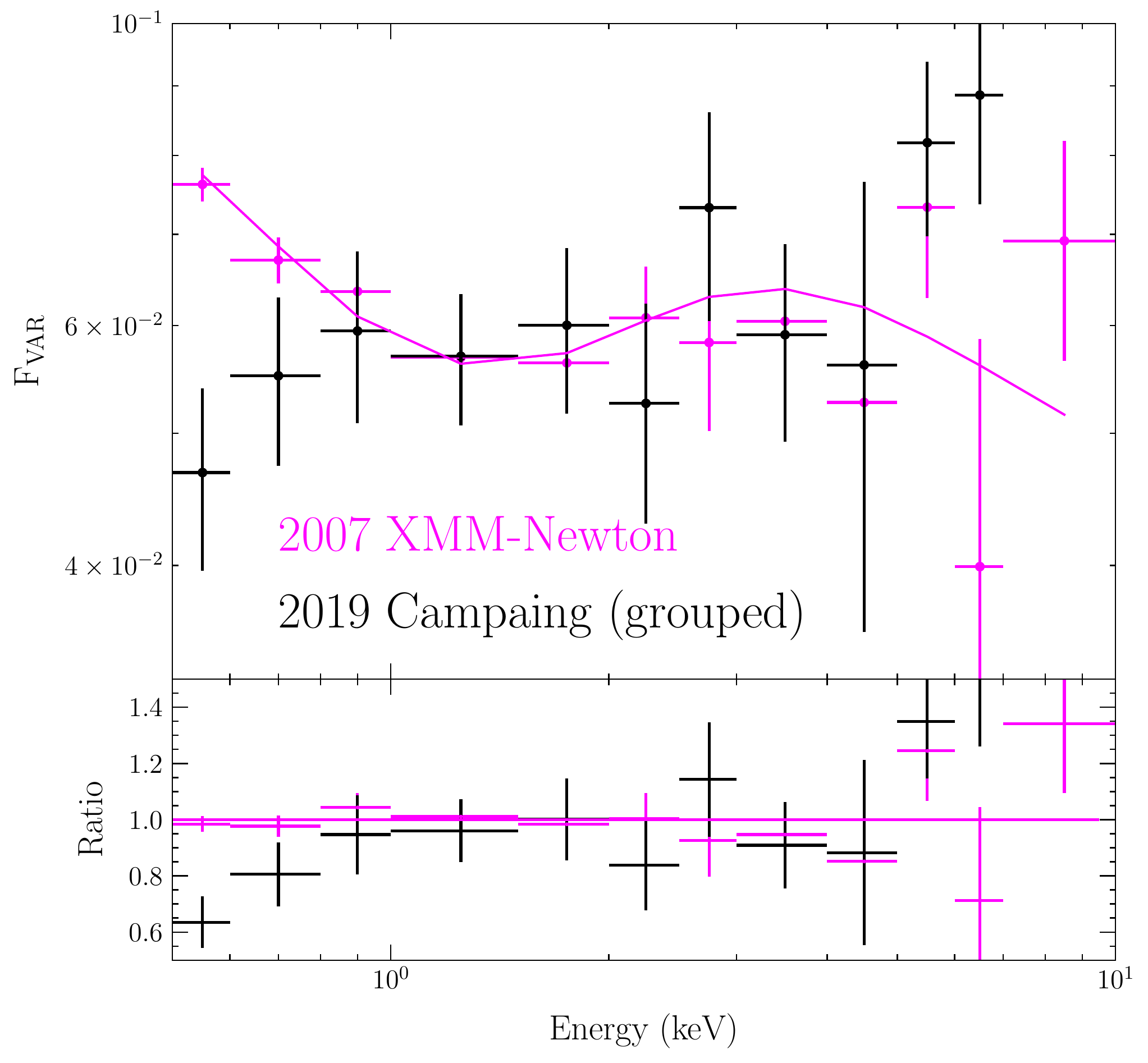}
	\caption{Comparison between the \small{F$_{\rm var}$ spectrum of the 2007 (magenta) and 2019 (black) observations respectively, where the latter was obtained by grouping together the 5 observations via the standard command \textsc{setplotgroup} within \textit{XSPEC}. The magenta curve represents a theoretical F$_{\rm var}$ spectrum which accounts for combined contribution from a variable power-law-like continuum and a variable soft-excess.\label{fvarfit}}}
\end{figure}
\section{Timing properties}
We extracted the background subtracted light curves for all the observations. In the top panels of Fig.~\ref{lc} we show the soft and hard X-ray light curves of the 2019 monitoring campaign time series while those of archival observations ( \textit{XMM-Newton} and \textit{Suzaku}) are shown in the bottom panels. From Fig.\ref{lc}, both short- and long- term variability characterises the ESO 511-G030 light curves. During 2019, ESO 511-G030 had a quite stable behaviour with both the soft and hard X-rays being fairly constant within each exposure and among the different pointings. Such a constancy, apart from a small fraction of the fifth exposure, can be observed in the ratios panels. This constant behaviour suggests that the balance between soft and hard X-rays did not change during the campaign. The flat shape of the 2019 hardness ratios can be quantitatively(qualitatively) compared with those computed from \textit{XMM-Newton}(Suzaku) archival exposures. The ratios between the 0.5-2 and 2-10 keV bands was more variable in the 2007 \textit{XMM-Newton} exposure with changes of about 25\%. Hardness ratios from \textit{Suzaku} show moderate variations within the same exposure and varied of $\sim$10\% on daily rather than monthly timescales.\\
\indent  Short-term X-ray variations are related to the intrinsic properties of the AGN such as its SMBH mass or its luminosity \citep[e.g.][]{Vaughan2003,Papadakis2004,McHardy2006}. \cite{Ponti12} computed the normalised excess variance of ESO 511-G030 using the 2007 \textit{XMM-Newton} light curves. Such a variability estimator allowed the authors to derive the black hole mass of ESO 511-G030 to be $\log$M$_{\rm BH}$=7.89$_{-0.20}^{+0.30}$ M$_\sun$.\\
\indent Another commonly adopted estimator suitable for X-ray variability characterisation is the fractional root mean square variability amplitude \citep[F$_{\rm var}$, e.g.][]{Edelson2002,Vaughan2003,Ponti2004}. The  F$_{\rm var}$ tool is the square root of the normalised excess variance and it has been widely used to characterise the variability properties of AGN in X-rays \citep[e.g.][]{,Vaughan2004,Ponti2006,Matzeu2016mnras,Matzeu2017,Alston2019,Parker2020,DeMarco2020,Igo2020,Middei2020}. 
We studied the variability properties of ESO 511-G030 by computing the F$_{\rm var}$ spectra for each of the 2019
\textit{XMM-Newton} observations. These spectra sample variability on timescales ranging between $\sim1-40$ ks. We used the background subtracted light curves extracted in different energy intervals and adopting a temporal bin of 1000 sec. Following the same procedure, we also computed the F$_{\rm var}$ spectrum of the 2007 \textit{XMM-Newton} observation. The latter, samples longer timescales ranging between $\sim1-120$ ks, and enables us to identify variable spectral components contributing to the time-averaged spectrum. The resulting F$_{\rm var}$ spectra of each observation are shown in Fig.~\ref{fvar}. The errors are computed using Eq. B2 of \citet{Vaughan2003} and account only for the uncertainty caused by Poisson noise.\\
\indent Aside from some excess towards the high energy region of the spectra (e.g., due to residual background variability), all the observations from the 2019 monitoring show a rather flat F$_{\rm var}$ spectrum, therefore implying a similar variability power across different energy bands. On the contrary, the 2007 F$_{\rm var}$ spectrum clearly shows a divergence, from the 2019, in the soft X-rays, consistent with the presence of an additional variability component.\\
\indent To better highlight this, in Fig.~\ref{fvarfit} we overlay the 2007 F$_{\rm var}$ spectrum and the grouped F$_{\rm var}$ spectra from the 2019 campaign. We used publicly available table models\footnote{\url{https://www.michaelparker.space/variance-models},  \citealt{Parker2020}} to describe the 2007 F$_{\rm var}$ spectrum in terms of combined contribution from flux variability of a power-law-like continuum and a soft-excess. While the model well reproduces the 2007 F$_{\rm var}$ spectrum and the high energy part of the 2019 data, it clearly overestimates the soft band part of the 2019 F$_{\rm var}$ spectrum. One possible explanation is the presence of an additional soft variability component in 2007 which is not present in the 2019 data.

\section{Spectral properties: the \textit{XMM-Newton/NuSTAR} 2019 campaign}
\begin{figure}[t]
	\centering
	\includegraphics[width=0.49\textwidth]{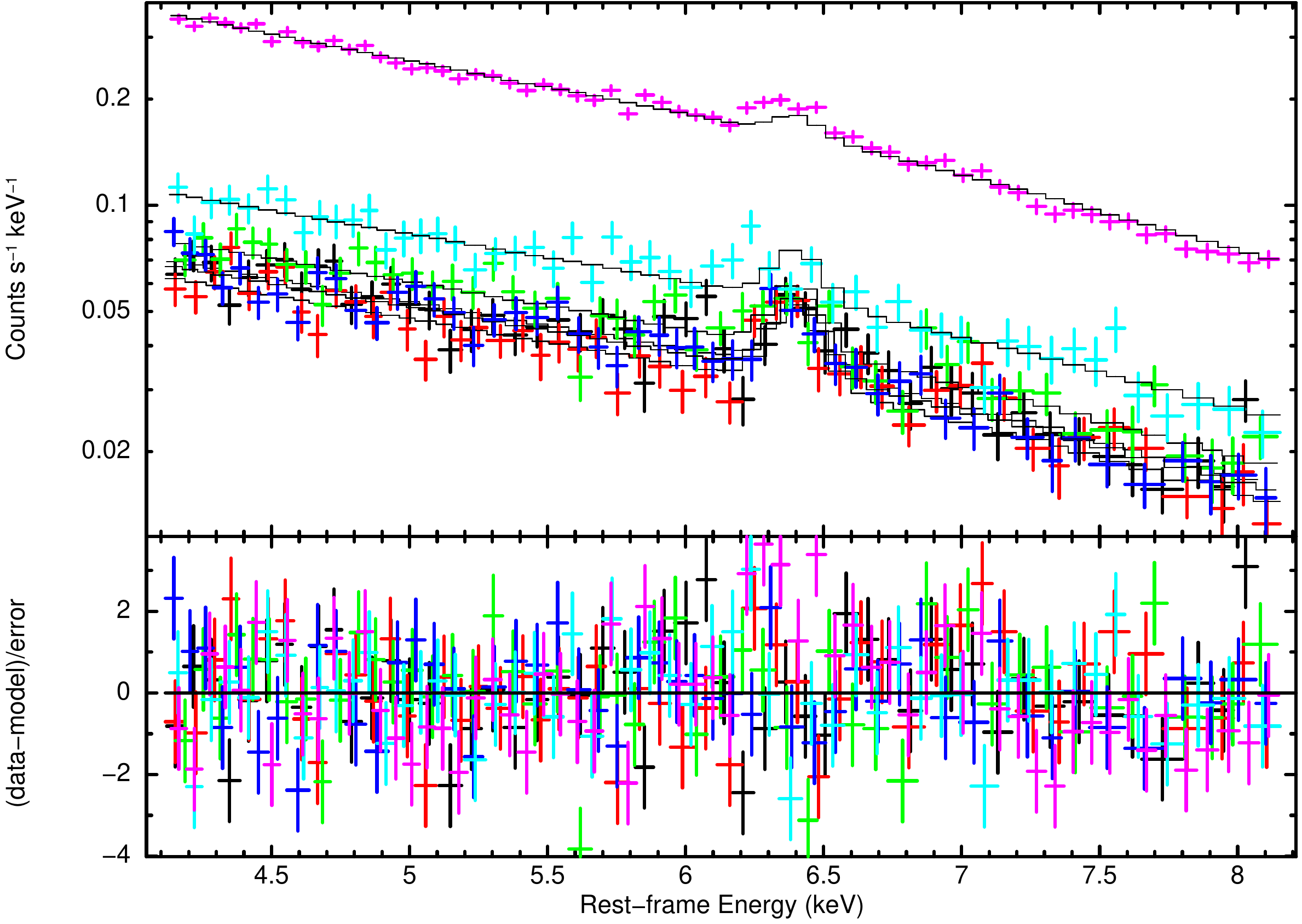}
	\includegraphics[width=0.49\textwidth]{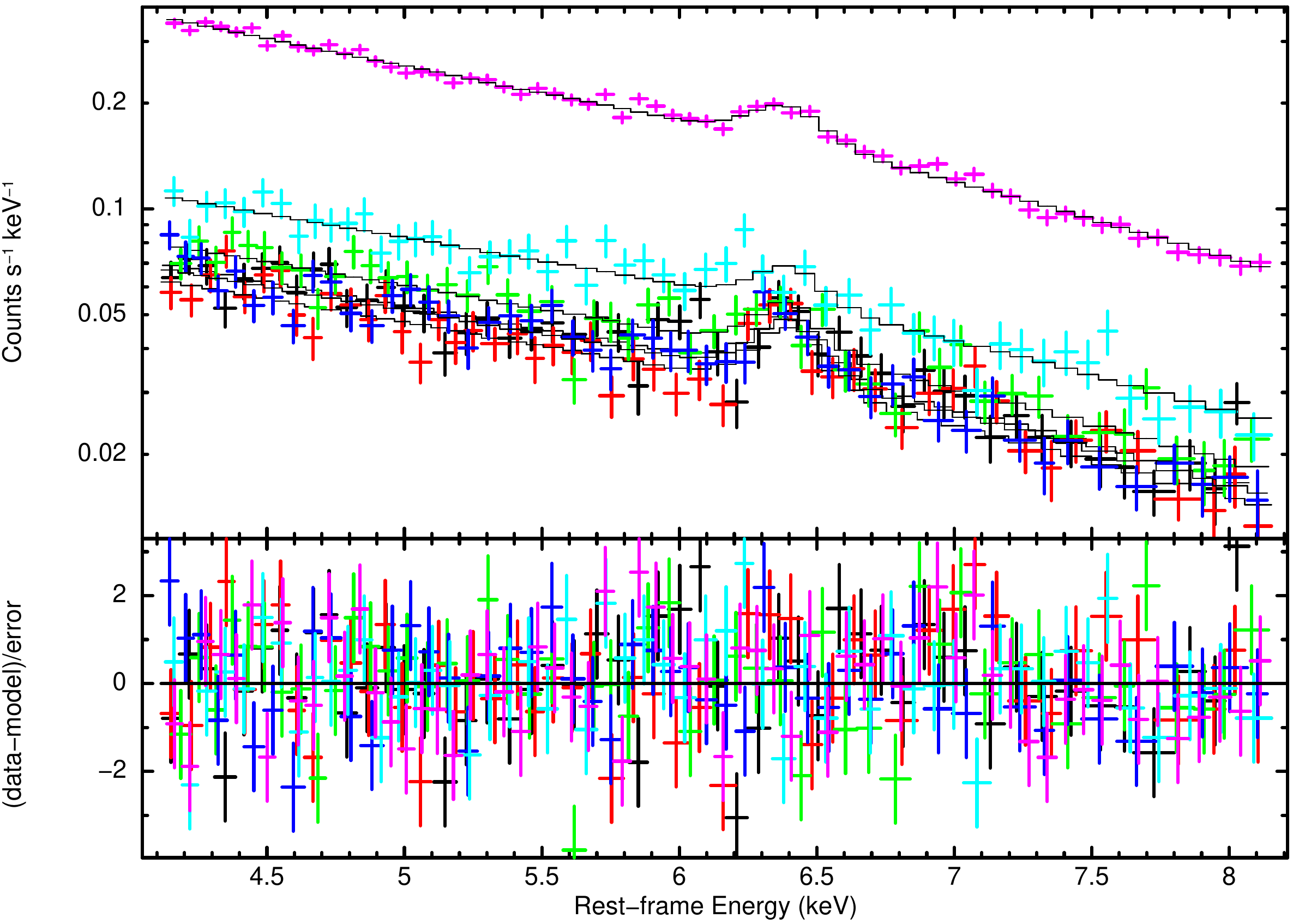}
	\caption{\small{Top panels: EPIC-pn data in the 4-8 keV energy band fitted using a power law and a narrow Gaussian emission line assumed to be constant among the exposures. Residuals in the Fe K energy range are present. Bottom panels: Same as above but with the addition of a broad Fe K$\alpha$ emission line for which the line normalisation was assumed to be constant among the 2019 pointings but free to vary in 2007.} \label{FeK}}
\end{figure}
\subsection{The Fe K$\alpha$ complex}
\begin{figure}[t]
	\centering
	\includegraphics[width=0.49\textwidth]{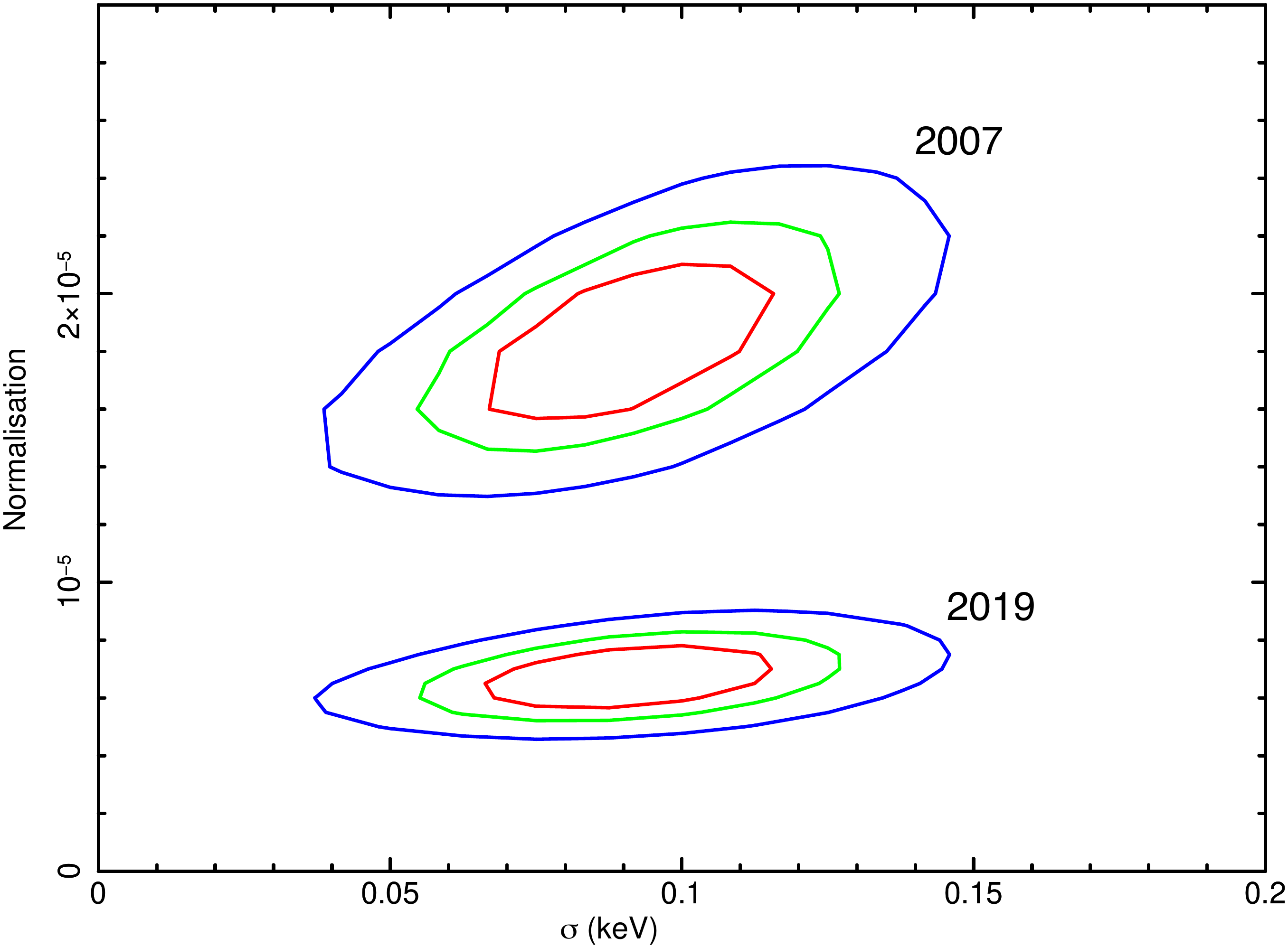}
   \caption{\small{Confidence regions at 99\%, 90\% and 68\% for the line's width and normalisation. These contours have been obtained assuming the model power-law+zGauss$_{\rm broad}$. Contours were obtained using a single broad Gaussian component.} \label{broad}}
\end{figure}
\indent We begun our investigation by focusing on 2019 \textit{EPIC-pn} spectra (between 4 and 8 keV) and the properties of the Fe K$\alpha$ emission line. We adopted two components, a power-law and Gaussian line assumed to be narrow ($\sigma$=0 eV) and centered at 6.4 keV. We simultaneously fitted all the spectra computing the photon index (we assumed its value to be the same among the pointings) and the power-law normalisation. For the Gaussian component, we  calculated its normalisation in all the exposures. This simple fit returned a Fe K$\alpha$ flux consistent with being constant, Norm$_{\rm Fe K\alpha}$= (6.8$\pm$0.8)$\times$10$^{-6}$ photons cm$^{-2}$ s$^{-1}$. We then considered the 2007 \textit{XMM-Newton} data on the 4-8 keV energy range on which we tested the same model and further assumed the Fe K$\alpha$ to have the same normalisation between 2007 and 2019. In other words, we only fitted the photon index and the normalisation of the continuum for this newly added data, see top panel Fig.~\ref{FeK}. However, the narrow emission line only reproduces the 2019 spectra and an additional broader component is required for the 2007 data. We thus froze the narrow Gaussian component to its best-fit value and added a new broad Gaussian emission line. In the fit, the line energy centroid and width were computed and we assumed these values to be the same among the all observations. We only allowed the line's normalisation to vary between the 2019 dataset and the 2007 exposure. This new model resulted in the fit of Fig.~\ref{FeK}, bottom panel, with $\chi^2$/d.o.f.=400/310. We subsequently re-fit the data also allowing the narrow Gaussian normalisation to vary between the datasets. This test resulted in a slight benefit in terms of statistic with $\Delta\chi^2$/$\Delta$d.o.f.=-10/-1. However, in this case, only an upper limit is returned for the narrow Gaussian components' flux.\\
\indent We further tested the origin of the Fe K$\alpha$ emission line computing the fit once again but only including a broad Gaussian component. Again we assumed the line's energy centroid and width to be the same across the years and only its normalisation was calculated separately for the 2019 and 2007 data. Interestingly, this step led to a statistically equivalent fit $\chi^2$/d.o.f.=400/310, which suggests the Fe K$\alpha$ in ESO 511-G030 to be consistent either with a superposition of a narrow and constant core plus a broad and variable component or with a single and moderately broad Gaussian that varies in time becoming stronger at a higher continuum flux. This is illustrated in Fig.~\ref{broad}, whereby the line normalisation for a single broadened Gaussian significantly decreased between the high flux 2007 and low flux 2019 observations. This behaviour, quite at odds with what commonly observed in AGNs, was also seen in NGC 2992 \citet{Marinucci2020}.

\subsection{Spectral modelling}

\indent We investigated the broadband spectral properties of ESO 511-G030 by testing a purely phenomenological model. We modelled the \textit{XMM-Newton/NuSTAR} data with a cut-off power-law absorbed for the Galaxy \citep[N$_{\rm H}$= 4.33$\times10^{20}$ cm$^{-2}$,][]{HI4PI}, a moderately broad Gaussian component for the Fe K$\alpha$, and a thermal component to account for curvature in the soft band (Model A). We fitted separately each observation and we report the inferred best-fit values and the statistic associated to each fit in Table~\ref{campaignT}. This procedure revealed that no significant spectral variations occurred during the campaign. The Fe K$\alpha$ was constant in terms of normalisation and in all but observations 2 and 4 the line profile was consistent with being broad. The high energy cut-off was constrained in observations 1 and 5 while only lower limits were obtained in the remaining exposures. At lower energies, a weak and constant black-boky-like component did not vary among the different observations. Given the little variability among the parameters, we fitted simultaneously all the observations tying the photon index, the cut-off energy for the primary continuum, the energy centroid and width of the Gaussian component and the temperature and normalisation of the black body. This resulted into a fit with $\chi^2$=1470 for 1416 d.o.f and an associated null probability of 0.1. Moreover we found a $\Gamma$=1.62$\pm$0.02 while the high energy roll over was as E$_{\rm c}$>160 keV. For the Fe K$\alpha$ we obtained E$_{\rm FeK\alpha}$=6.39$\pm$0.02, $\sigma_{\rm FeK\alpha}$=70$\pm30$ and a normalisation of Norm$_{\rm FeK\alpha}$=7.2$\pm1.8\times10^{-6}$ photons cm$^{-2}$ s$^{-1}$, in full agreement with the results reported in Sect 4.1.

\begin{figure}
		\centering
		\includegraphics[width=0.49\textwidth]{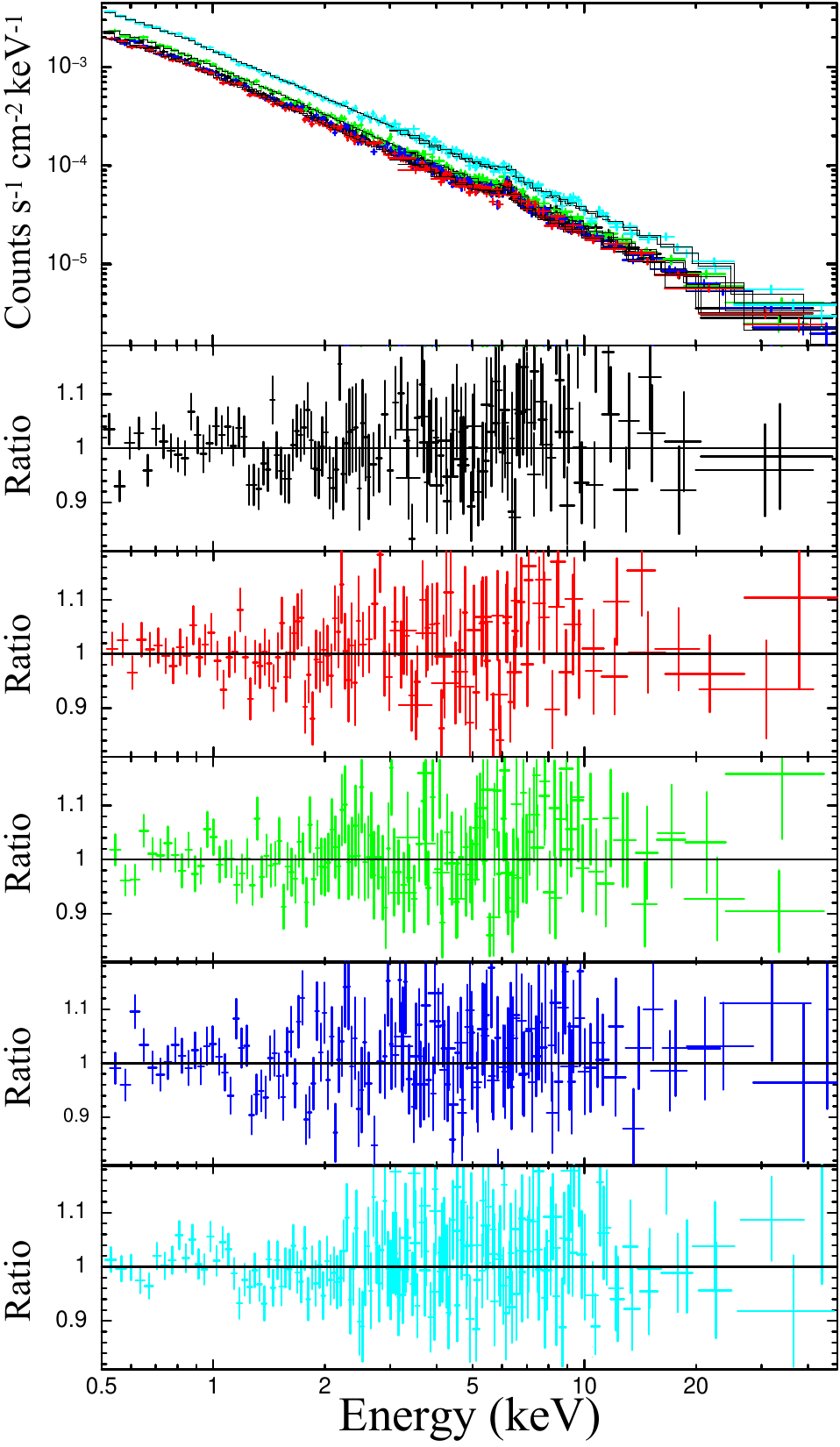}
		\caption{\small{{Model C fitting the to the 2019 spectra. \textit{Borus} and \textit{relxill} account for the cold and hot reflection components, respectively.}} \label{borusrelxill}}
\end{figure}

\begin{table*}
	\centering
	\setlength{\tabcolsep}{2.5pt}
	\caption{\small{Best-fit parameters obtained using models A and B (see Sect. 4.2 for details). The corresponding statistics are also showed. For the sake of simplicity, we only quote the constant accounting for the  \textit{XMM-Newton} spectra as those for  \textit{FPMA\&B} are always within a 10\% it.	Finally, the L39 quantity refers to the luminosity in units of 10$^{39}$ erg s$^{-1}$ and D$_{\rm 10}$ is the distance in units of 10 kpc, as defined in the XSPEC manual (see, https://heasarc.gsfc.nasa.gov/xanadu/xspec/manual/node137.html}.}\label{campaignT}
	\begin{tabular}{c c c c c c c c c}	
		Model & Comp. & Par. & Obs. 1  & Obs. 2 & Obs. 3 & Obs. 4 & Obs. 5 & Units\\
		\hline
		Model A&&&&&&&&\\
		&blackbody& kT&0.15$\pm$0.02 & 0.15$\pm$0.01&0.15$\pm$0.02&0.15$\pm$0.02&0.14$\pm$0.01&eV\\
		&& Norm&6.0$\pm$0.8&5.2$\pm$0.8&5.8$\pm$0.7&4.8$\pm$1.1&9.5$\pm$1.2&$\times10^{-6}$ L39\\
		&cut-off pl& $\Gamma$&1.64$\pm$0.02&1.64$\pm$0.02&1.63$\pm$0.03&1.65$\pm$0.01&1.61$\pm$0.01&\\
		&&E$_{\rm cut}$&160$^{+175}_{-75}$&>100&>345&>100&150$^{+100}_{-50}$&keV\\
  &&Norm&9.4$\pm$0.1&8.4$\pm$0.1&10.8$\pm$0.1&9.0$\pm$0.2&15.5$\pm$0.2&$\times10^{-4}$ ph keV$^{-1}$ cm$^{-2}$ s$^{-1}$\\ 
    &zGauss&E&6.42$\pm$0.09&6.38$\pm$0.03&6.39$\pm$0.12&6.40$\pm$0.04&6.38$\pm$0.10&keV\\
		&&$\sigma$&100$^{+140}_{-90}$&<65&290$\pm$75&<120&170$^{+200}_{-100}$&eV\\
		&&Norm&7.6$\pm$2.2&8.6$\pm1.3$&8.8$\pm$2.0&8.3$\pm$&9.7$\pm$3.3&$\times10^{-6}$ ph cm$^{-2}$ s$^{-1}$\\
  &$\chi^2$/d.of.&&290/260&280/256&310/266&320/287&350/331\\
  &p.null&&0.1&0.1&0.05&0.1&0.2\\

		\hline
		Model C &&&&&&&&\\
        
        &cutoffpl&$\Gamma$&1.73$\pm$0.03&1.78$\pm$0.05&1.71$\pm$0.04&1.75$\pm$0.03&1.69$\pm$0.02&\\
        &&E$_{\rm cut}$&>170&>70&>80&>135&75$\pm$20&keV\\
        &&Norm&1.02$\pm$0.01&0.93$\pm$0.01&1.10$\pm$0.01&0.98$\pm$0.01&1.70$\pm$0.01&$\times10^{-3}$ ph keV$^{-1}$ cm$^{-2}$ s$^{-1}$\\
 	    
        &\textit{relxill}& R$_{\rm in}$&>15&-&-&>20&>15&R$_{\rm g}$\\
        &&$\log \xi$ &1.5$\pm$0.4&>1.1&2.0$\pm$0.6&1.7$\pm$0.5&1.3$\pm$2& \\  
        &&Norm&3.0$\pm$0.5&2.2$\pm$0.3&2.8$\pm$1.2&2.8$\pm$1.0& 6.5$\pm1.1$&$\times10^{-6}$ ph keV$^{-1}$ cm$^{-2}$ s$^{-1}$\\
		
  &\textit{Borus}& Norm &1.2$\pm$0.8&2.0$\pm$0.8&0.97$\pm$0.23&1.8$\pm$0.5&0.80$\pm$0.40&$\times10^{-3}$ ph keV$^{-1}$ cm$^{-2}$ s$^{-1}$\\ 
		&& $\rm \log N_{ H}$&23.3$\pm$0.2&23.2$\pm$0.2&23.5$\pm$0.2&23.2$\pm$0.1&23.1$\pm$0.2&1/(cm$^{-2}$)\\
       
       &$\chi^2$/d.of.&&280/260&282/256&296/266&320/287&370/331\\
        &p.null&&0.2&0.13&0.1&0.1&0.1\\
  
		\hline
        &Flux$_{\rm 0.5-2~keV}$&&2.1$\pm$0.1&1.85$\pm$0.15&2.4$\pm$0.2&2.0$\pm$0.1&$3.2\pm$0.1&$\times10^{-12}$ erg cm$^{-2}$ s$^{-1}$\\
		&Flux$_{\rm 2-10~keV}$&&4.0$\pm$0.1&3.7$\pm$0.1&4.8$\pm$0.1&4.0$\pm$0.2&6.5$\pm$0.1&$\times10^{-12}$ erg cm$^{-2}$ s$^{-1}$\\
	\end{tabular}
\end{table*}

\indent We subsequently tested a more reliable physical framework for the ESO 511-G030 2019 spectra. We started considering two scenarios: (i) one accounting for a narrow Fe K$\alpha$, signature of distant reflecting material, and (ii) in which this emission feature is a blend of a relativistically broadened and the narrow components. The model \textit{Borus} \citep[e.g.][]{Balokovic2018} was used to account for the distant reflection and \textit{relxill} \citep[e.g.][]{Garc13,Garc14} for the relativistic component. Within \textit{Borus}, the toroidal X-ray reprocessor is assumed to have a spherical shape with conical cutouts at both poles and the X-ray source is assumed to be at its center. We used the table {\it borus01\_v161215a.ftz}.
\textit{Relxill} \citep[e.g.][]{Daus16} is part of a model suite accounting for ionised reflection from an accretion disc illuminated by a hot corona. In \textsc{XSPEC} notation we thus tested the following models: $\rm tbabs_{\rm G} \times (cutoffpl+Borus)$, Model B, and $\rm tbabs_{\rm G} \times (cutoffpl+relxill+Borus)$, Model C, for cases (i) and (ii), respectively.
These models were applied to each {\it XMM-Newton/NuSTAR} dataset and we fitted the $\Gamma$, the high energy cut-off and the normalisation of the primary continuum tying these values with those of the \textit{Borus} table. The column density and the normalisation of the \textit{Borus} table were also computed in each exposure. We proceeded similarly when testing Model C. In this case, we assumed the Iron abundance to be solar (A$_{\rm Fe}$=1) and computed the ionisation parameter $\xi$ and the inner radius r$_{\rm in}$. Model C better reproduces the data and Fig.~\ref{borusrelxill} reports the corresponding best-fit values.\\
The ESO 511-G030 spectra are well described by a primary continuum with $\Gamma$=1.73$\pm$0.02. Lower-limits for the high energy cut-off were inferred in all but observation 5 for which E$_{\rm c}$=75$\pm$20 keV was obtained. The narrow core of the Fe K$\alpha$ emerges from a Compton-thin medium with an averaged column density N$_{\rm H}\sim1.8\times$10$^{23}$ cm$^{-2}$ which is also responsible for the moderate high energy curvature of the spectra. A relativistic reflection component is likely originating from mildly ionised matter with the ionisation parameter being consistent among the exposures. However, the inner radius of this component is poorly constrained. The best-fit quantities inferred from the fits are quoted in Table~\ref{campaignT}.\\
\indent As a final test, we added a black body component to account for any weak underlying soft X-ray spectral feature in each observation. The addition of this soft-component did not provide any significant improvement to the fit. This test is in agreement with a scenario of a absent/weak soft X-ray excess in this source.

\section{Archival X-ray observations of ESO 511-G030}
\begin{figure}
		\centering
		\includegraphics[width=0.49\textwidth]{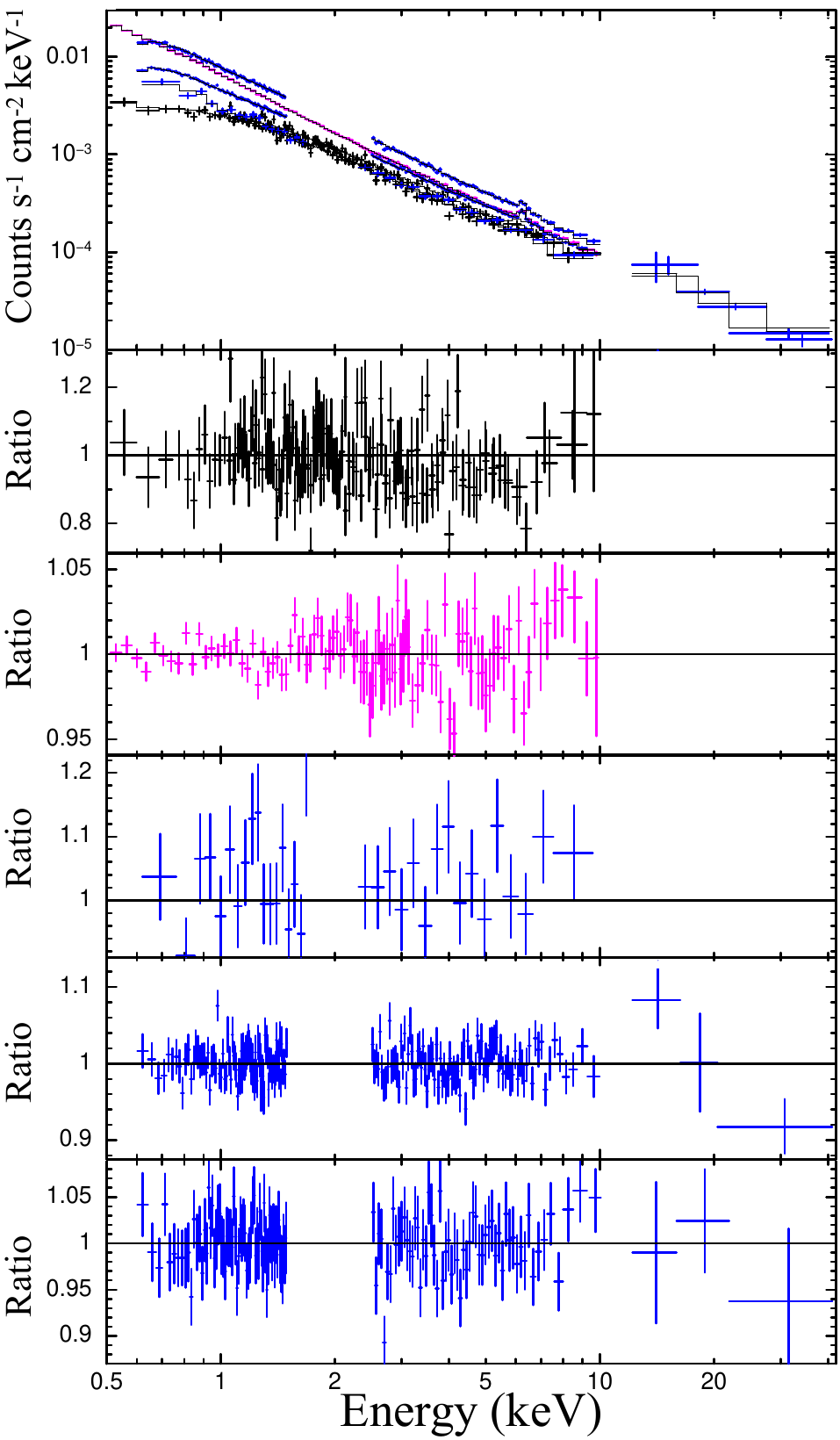}
		\caption{\small{Fit to the \textit{ASCA} (black), 2007 \textit{XMM-Newton} (magenta) and \textit{Suzaku} (blue) data. Bottom panels account for the data-model ratios.}\label{archivefit}}
\end{figure}

Hitherto, ESO 511-G030 AGN has been observed by different facilities. Detailed studies on \textit{ASCA}, \textit{Suzaku} as well as the 2007 \textit{XMM-Newton} exposure have already been published \citep[e.g.][]{Turner2001asca,Ghosh2020}, thus we will perform straight forward fits on these data to extract information on the spectral properties of ESO 511-G030 in previous years.\\
\indent We tested our Model C, adopting the same fitting  procedure described in previous Sect 4.3 and, slightly modifying the model when needed. \\
\indent As clearly shown in Fig.~\ref{initialfig}, all archival exposures have larger fluxes than the data in 2019 and different spectral shapes, especially in the soft band. For this reason, we fitted separately \textit{ASCA}, all the \textit{SUZAKU} observations and \textit{XMM-Newton} data.  Concerning \textit{SUZAKU}, the additional cross-calibration constant (\textit{k}) was used for the \textit{XIS} and \textit{PIN} data.\\
Interestingly, our baseline Model C failed in reproducing all the archival spectra and additional components to this model are needed. In particular, a neutral absorption was required by
\textit{ASCA} data for which no relativistic reflection was necessary. For the \textit{SUZAKU} data, instead, we added a single black body (\textit{bb}) component to account for the soft-excess observed below $\sim$1 keV, while two \textit{bb}s were added to fit the 2007 \textit{XMM-Newton} exposure \citep[see also][]{Ghosh2020}. These steps yielded the fits in Fig.~\ref{archivefit} and in Table~\ref{archivetable} we report the corresponding best-fit values.
\begin{table*}
	\centering
	\setlength{\tabcolsep}{10.5pt}
	\caption{\small{Parameters derived adapting the baseline Model C on the archival data. The L39 stands for the luminosity in units of 10$^{39}$ ergs s$^{-1}$ and D$_{10}$ is the distance in units of 10 kpc}.}\label{archivetable}
	\begin{tabular}{c c c c c c c }	
		Mission & Comp. & Par. & Visit 1 & Visit 2 & Visit 3 &  Units\\
		\hline
        \textit{ASCA}&tbabs&N$_{\rm H}$ &1.0$\pm$0.2 &  & &$\times$10$^{21}$ cm$^{-2}$\\
		&\textit{Borus}& $ \log \rm N_{ H}$&>23.1&& &\\
        &&Norm &1.6$\pm$1.5&& &$\times10^{-2}$ ph keV$^{-1}$  cm$^{-2}$ s$^{-1}$\\ 
        &power-law& $\Gamma$&1.91$\pm$0.08&& &\\
		&&E$_{\rm c}\dagger$&500 && &keV\\
        && Norm&3.9$\pm$0.3&& &$\times10^{-3}$ ph keV$^{-1}$  cm$^{-2}$ s$^{-1}$\\ 
        &$\chi^2$/d.o.f.&&360/403 & & &\\
		&p.null&&0.9 & & &\\
		\hline
		&Flux$_{\rm 0.5-2~keV}$& &6.0$\pm$0.1 & & &$\times10^{-12}$ erg cm$^{-2}$ s$^{-1}$\\
		&Flux$_{\rm 2-10~keV}$&&14.0$\pm$0.5& & &$\times10^{-12}$ erg cm$^{-2}$ s$^{-1}$\\
		\hline
		\hline
		\\
		\textit{XMM-Newton}&bb& T$_{\rm bb}$ &180$\pm$15 & & & eV\\
		&&Norm &3.2$\pm$0.4& & &$\times$10$^{-5}$ L39\\  
		&bb& T$_{\rm bb}$ &75$\pm$10& & &eV\\
		&&Norm &1.0$\pm$0.2& & &$\times$10$^{-4}$ L39\\  
		&\textit{relxill}&r$_{\rm in}$&25$^{+12}_{-20}$&&&R$_{\rm g}$\\
		& &$\log\xi$&1.5$\pm$0.2&&&\\
		& &Norm& 1.8$\pm$0.6&&&$\times$10$^{-5}$ ph keV$^{-1}$ cm$^{-2}$ s$^{-1}$\\
		&\textit{Borus}& $\log \rm N_{ H}$&23.8$\pm$0.3&& &\\
		&&Norm &3.0$\pm$0.6& & &$\times10^{-3}$ ph keV$^{-1}$  cm$^{-2}$ s$^{-1}$\\
        &power-law& $\Gamma$&1.88$\pm$0.02&& &\\
		&& Norm&6.5$\pm$0.1&& &$\times10^{-3}$ ph keV$^{-1}$  cm$^{-2}$ s$^{-1}$\\
  		&$\chi^2$/d.o.f.&&180/160 & & &\\
  		&p.null&&0.2 & & &\\
		\hline
		&Flux$_{\rm 0.5-2~keV}$&&15.0$\pm$0.5 & & &$\times10^{-12}$ erg cm$^{-2}$ s$^{-1}$\\
		&Flux$_{\rm 2-10~keV}$&&20.0$\pm$0.5& & &$\times10^{-12}$ erg cm$^{-2}$ s$^{-1}$\\
		\hline
		\hline
		\\
		\textit{SUZAKU}&bb& T$_{\rm bb}$& 84$\pm$5&85$\pm$6&115$\pm$15 &eV\\
		&&Norm &1.3$\pm$0.8 &1.4$\pm$0.3&1.1$\pm$0.3 &$\times$10$^{-4}$ L39\\  
		&\textit{relxill}&r$_{\rm in}$&-&>2&>10&R$_{\rm g}$\\
		& &$\log\xi$&>1.2&2.7$\pm$0.2 &2.9$\pm$0.2&\\
		& &Norm& 0.85$\pm$55&0.75$\pm$0.35&2.1$\pm$1.2&$\times10^{-5}$ ph keV$^{-1}$  cm$^{-2}$ s$^{-1}$\\
		&\textit{Borus}& $\rm  \log N_{\rm H}$&22.7$\pm$0.4&23.7$\pm$0.2&23.4$\pm$0.5 &\\
      	&&Norm &2.0$\pm$1.0 &2.1$\pm$0.4&2.3$\pm$0.4 &$\times10^{-2}$ ph keV$^{-1}$  cm$^{-2}$ s$^{-1}$\\ 
      	
		&power-law& $\Gamma$&1.77$\pm$0.02&1.80$\pm$0.02&1.85$\pm$0.02 &\\
		&& E$_{\rm c}$&500$\dagger$&>120&$>$450 &keV\\
		&& Norm&3.35$\pm$0.4&5.0$\pm$0.1&7.4$\pm$0.3 &$\times10^{-3}$ ph keV$^{-1}$  cm$^{-2}$ s$^{-1}$\\ 
		&const& k &-&1.3$\pm$0.10&1.16$\pm$0.08 &\\ 
  		&$\chi^2$/d.o.f.&&312/306 &1527/1437 & 1017/967&\\
  		&p.null&&0.42 &0.04&0.097 &\\
		\hline
		&Flux$_{\rm 0.5-2~keV}$&&8.2$\pm$0.4 &11.7$\pm$0.3 &21.0$\pm$0.7 &$\times10^{-12}$ erg cm$^{-2}$ s$^{-1}$\\
		&Flux$_{\rm 2-10~keV}$&&13.0$\pm$0.5&19.1$\pm$0.3&27.7$\pm$0.4& $\times10^{-12}$ erg cm$^{-2}$ s$^{-1}$\\
		\hline
		\hline
	\end{tabular}
\end{table*}

Apart from the 1999 absorption event observed in the {\it ASCA} data due to matter with constant N$_{\rm H}$=1.2$\pm$0.3 $\times$10$^{21}$ cm$^{-2}$, ESO 511-G030 archival spectra are consistent with a variable power-law that is, on average, softer than in 2019 ($\Delta\Gamma\sim0.2$). The continuum flux in the 2-10 keV energy range is up to a factor of $\sim$6 larger than in 2019 while the soft X-ray flux in the 0.5-2 keV range is even larger, up to a factor of $\sim$10. The large changes in the soft flux can be ascribed to the presence(lack) of the soft-excess component. The soft-excess clearly plays a major role in shaping the soft band in the 2007 \textit{XMM-Newton} and 2011 \textit{SUZAKU} observations.\\
\indent The relativistic reflection observed in these archival observations is about $\sim$10 times larger than the one inferred in 2019, in agreement with the hot reflection responding to the primary changes. On the other hand, the reflected spectrum due to cold matter and also contributing to the Fe K$\alpha$ has, has a more constant behaviour across the years. Values derived for the normalisation the \textit{BorusK/L} tables are, in fact, rather constant among the epochs. Thus, to a lower(higher) primary flux would correspond a larger(smaller) reflection fraction. Finally, the Compton-thin nature of the reprocessor in ESO 511-G030 is further confirmed by this data.

\section{The {\it Swift} Monitoring campaign}
\begin{figure}[!]
	\centering
	\includegraphics[width=0.49\textwidth]{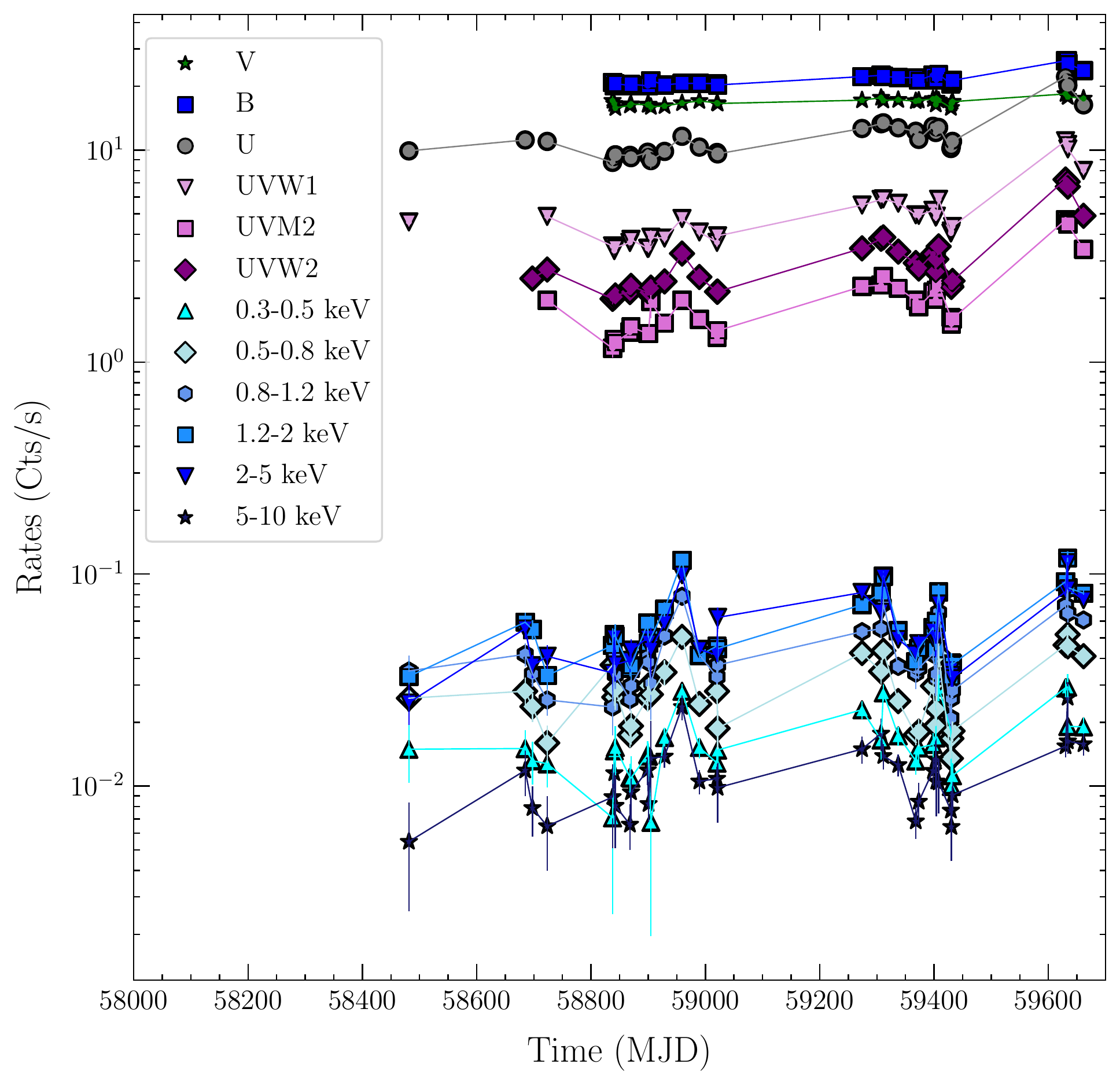}
	\caption{\small{Multi filter light curves derived from 2018 up to 2022 for ESO 511-G030 using\textit{Swift} -XRT and -UVOT. Until the end of 2021 (MJD<59580), after which a moderate increase of the ultraviolet emission can be observed, the optical-UV curves can be highly affected by the host emission. We notice that to quantify the actual amplitude of the optical-UV emission an estimate of the contribution of the host galaxy is required (see Sec 6.1). }} \label{swiftlc}
\end{figure}
\begin{figure*}[!]
	\centering
	\includegraphics[width=0.99\textwidth]{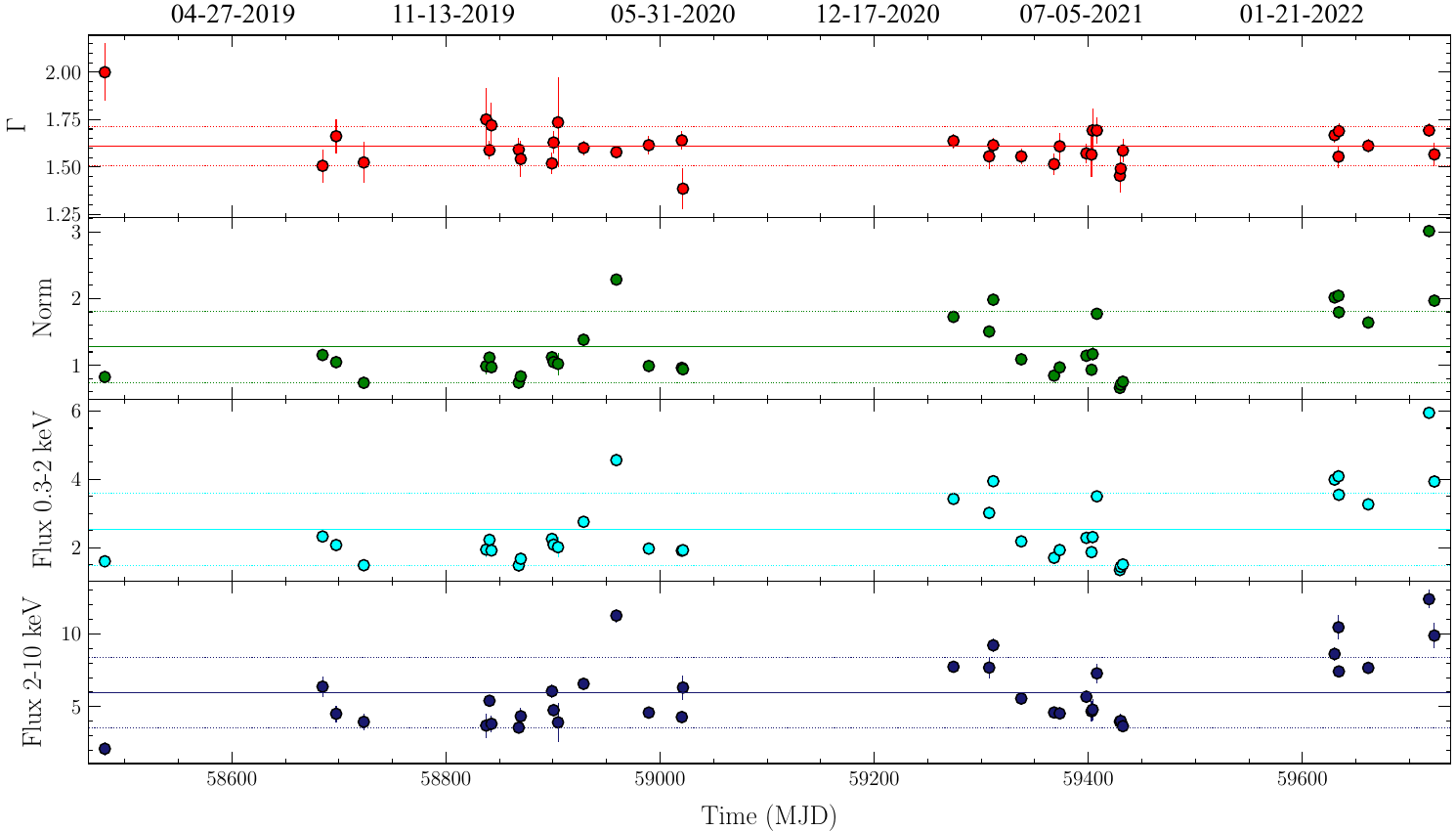}
	\caption{\small{Best fit parameters and fluxes in the 0.3-2 (Soft) and 2-10  keV (hard) bands derived from \textit{Swift}. Straight solid lines account for the mean value of the parameter while dotted lines represent the corresponding standard deviation. Fluxes are in units of $\rm 10^{-12} erg~cm^{-2}~s^{-1}$. \label{fitswift}}}
\end{figure*}
\begin{figure}[t]
	\centering
	\includegraphics[width=0.49\textwidth]{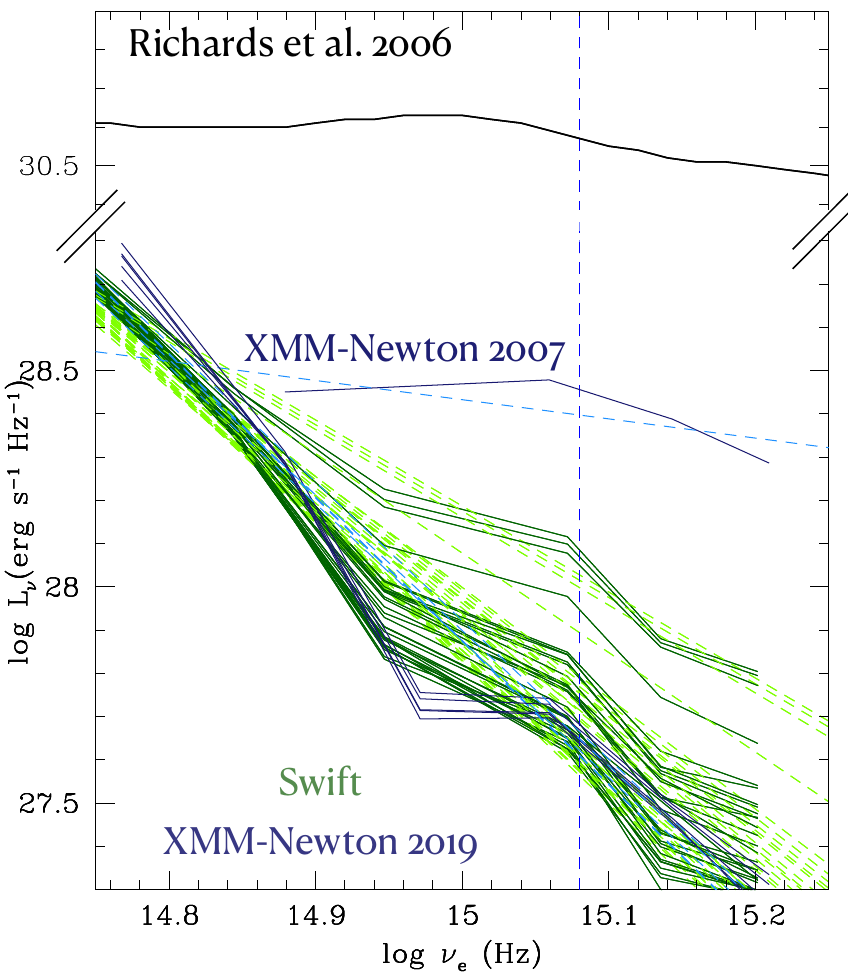}
	\caption{\small{ESO 511-G030 optical/UV SEDs derived using {\it XMM-Newton} (blue) and Swift data (dark green). The corresponding least squares fits are showed as straight dashed lines. The solid thick black line is the average SED by \citet{Richards2006} for type-1 objects in the SDSS. The vertical dashed line is at $\log\nu_*=15.08$, corresponding to 2500\AA. We notice the remarkable change in spectral shape from the 2007 observation versus the later ones.}\label{SED}}
\end{figure}
\indent We present here \textit{Swift} X-ray and ultraviolet data that were taken in the context of two different observational campaigns, one of which is still ongoing, performed between 2020 and the present year. The objective of the campaigns is to keep track of the broadband properties of ESO 511-G030 and, possibly, observe a revenant soft-excess. Up to Fall 2021, however, the source flux state has been consistent with the one observed on 2019 with \textit{XMM-Newton} and \textit{NuSTAR}. In Fig.~\ref{swiftlc}, we show the light curves derived for different X-ray bands (0.3-0.5, 0.5-0.8, 0.8-1.2, 1.2-2, 2-5 and 5-10 keV) and optical-UV filters (which are possibly dominated by the host galaxy). The {\it Swift-XRT} data show a min-to-max variability of about a factor of 5 across light curve. Once converted into fluxes, both X-rays and UVs are about a factor of $\sim$10 fainter than in the \textit{XMM-Newton} observation of 2007, thus ESO 511-G030 as observed with \textit{Swift} appears to be an extension of the quiescent state of ESO 511-G030 observed during 2019. By analysing the {\it Swift} spectra, our main objective was to establish whether a simple power-law component (plus Galactic absorption) was enough to explain the data. Our test, confirmed that a simple power-law well represents the 0.3-10 keV energy range of our source and no additional components are required. In Table~\ref{tableswift} the inferred best-fit quantities are listed and the corresponding  best-fit values are showed as a function of the observing time in Fig.~\ref{fitswift}. \textit{Swift} data are consistent with a fairly flat spectral shape of $\Gamma$=1.62$\pm$0.09 and only small variability is observed in the power-law normalisation. This may suggest the source still remains in a quiescent state with no soft excess awakening.\\
\indent To test whether the soft-excess was or not present in {\it XRT} spectra, we used all the observations quoted in Table~A.1 to produce the stacked spectrum of ESO 511-G030. The obtained spectrum has no signature of this component. A simple power-law absorbed by the Galaxy ($\Gamma$=1.62$\pm$0.02) is in fact enough to account for the data ($\chi^2$=337 for  339 d.o.f.),  this ruling out any additional component. The Swift fluxes in the 0.5-2 and 2-10 keV bands are compatible with what was observed during the 2019 {\it XMM-Newton/NuSTAR} monitoring campaign with F$_{\rm 0.5-2~keV}$=(2.18$\pm$0.02)$\times10^{-12}$ erg cm$^{-2}$ s$^{-1}$and F$_{\rm 2-10~keV}$=(5.1$\pm$0.1)$\times10^{-12}$  erg cm$^{-2}$ s$^{-1}$. Moreover, the stacked spectrum does not show any evidence of a Fe K$\alpha$ emission line. The lack of this feature can be likely explained by the coupled effects of the source low flux and the small effective area of the {\it Swift-XRT} telescope in the hard X-rays.\\
\indent Finally, we notice that the last four {\it Swift} exposures of ESO 511-G030 were consistent with a flux increase of the source in both the X-ray and ultraviolet energy bands.

\section{Relation between X-rays and UVs}
\begin{figure*}[h]
	\centering
	\includegraphics[width=0.99\textwidth]{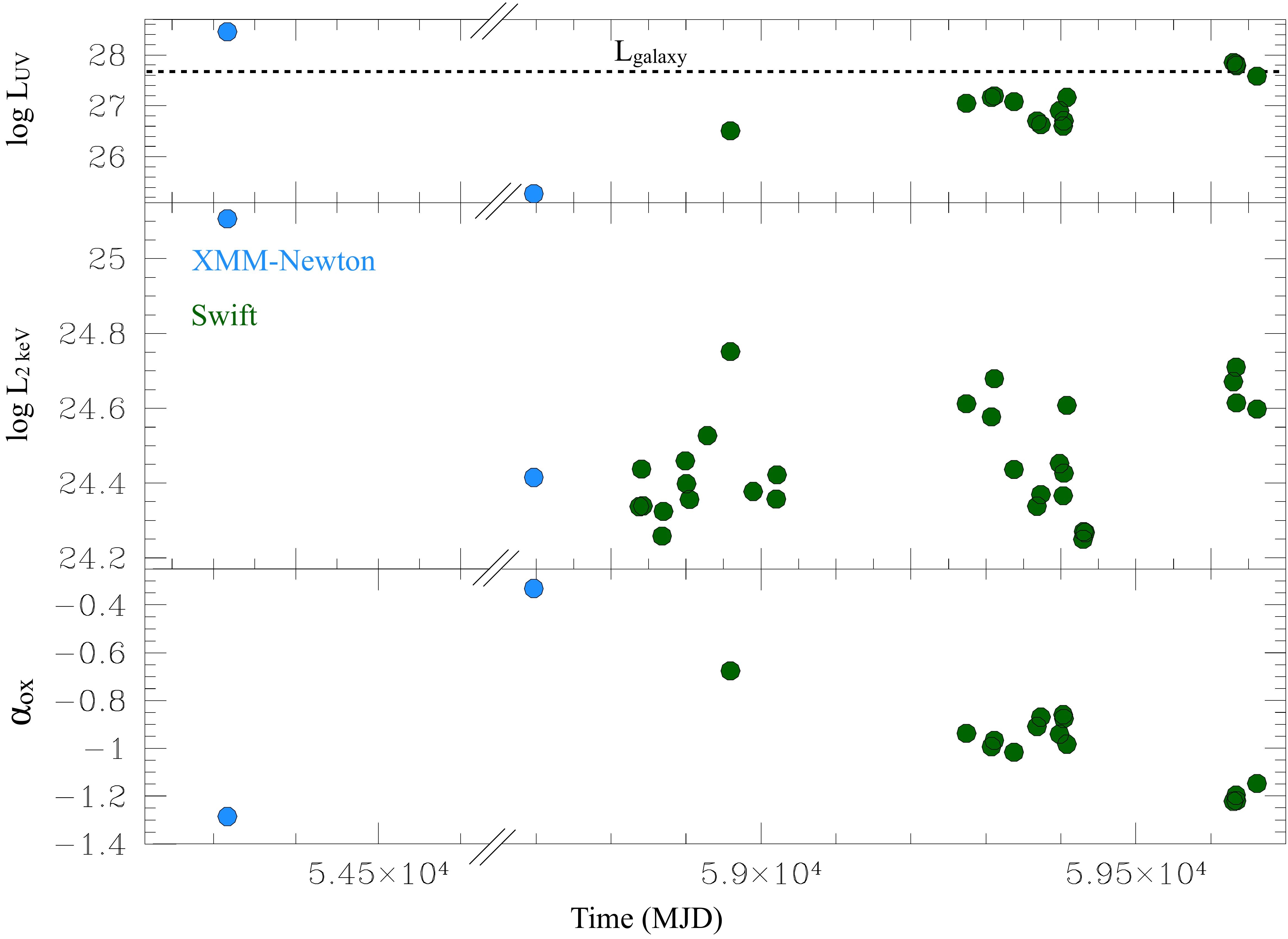}
	\caption{\small{Light curves for the AGN luminosity at 2500 \AA, at 2 keV, and the inferred $\alpha_{\rm OX}$. Blue dots identify values derived using \textit{XMM-Newton} observations while in green the same quantities were estimated using \textit{Swift}. The dashed horizontal line accounts for the adopted host galaxy luminosity at 2500 \AA~. Finally, both L$_{\rm UV}$ and L$_{\rm 2~keV}$ are in units of erg s$^{-1}$ Hz$^{-1}$.}\label{aox}}
\end{figure*}

A viable way to quantify the actual relation between X-rays and UVs is provided by the $\alpha_{OX}$ parameter. The non-linear relation between the UV and X-ray luminosity in AGNs was discovered in late 1970s by \citet[][]{Tananbaum1979} and several other authors investigating into the physical meaning of such a relation \citep[e.g.][]{Tananbaum1986,Zamorani1981,Vagnetti2010,Martocchia2017,Chiaraluce2018} and its implication in a cosmological framework \citep[e.g.][]{Risaliti2015,Lusso2016,Lusso2017,Lusso2020,Bisogni2021}.
Optical and ultraviolet data simultaneous with X-ray information are available for both {\it XMM-Newton} and {\it Swift}. For the 2007 {\it XMM-Newton} observation, only four OM filters are available (B, UVW1, UVM2, UVW2) while all of them were used in the 2019 monitoring campaign (V, B, U, UVW1, UVM2 and UVW2 at 5235\AA, 4050\AA, 3275\AA, 2675\AA, 2205\AA, 1894\AA, respectively). {\it Swift-XRT} observations are accompanied by one to six UVOT filters (V, B, U, UVW1 , UVM2 , UVW2  at 5468\AA, 4392\AA, 3465\AA, 2600\AA, 2246\AA and 1928\AA, respectively), this providing rich information on the UV continuum slope. We then derived the X-ray luminosity at 2 keV in order to compute the $\alpha_{OX}$ values. We relied on our best-fit to the EPIC-pn spectra discussed in Sect. 4 and 5 while the {\it XRT} spectra were modelled with a simple power-law with Galactic absorption, see Appendix A. 

\begin{figure}[ht]
	\includegraphics[width=0.49\textwidth]{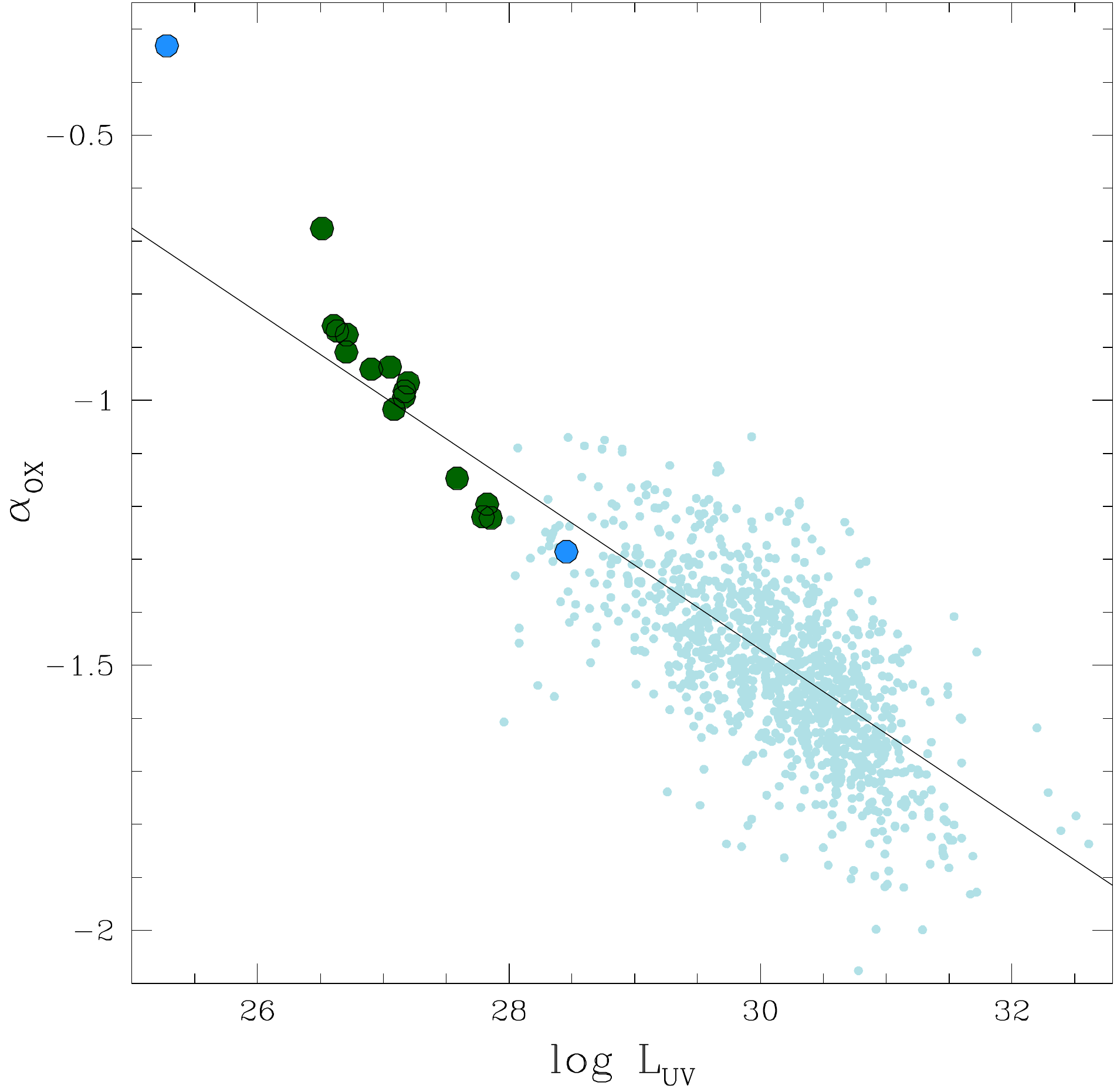}
	\caption{\small{The ESO 511-G030 $\alpha_{\rm OX}$ estimates comparing them with the sample by \citet[][]{Chiaraluce2018}.  The solid black line represents ${\rm logL_{\rm UV}-\alpha_{ox}}$ linear relation as derived in the same work.The $\alpha_{\rm OX}$ computed of ESO 511-G030 shows significant variability in agreement with the significant drop in the OM data showed in Fig.~\ref{initialfig}.}\label{aox1}}
\end{figure}

The total (AGN+galaxy) luminosity at $\log\nu_*=15.08$, corresponding to $\lambda$ = 2500\AA,  was estimated at each epoch as an interpolation between the data in the closest filters, which for these observations are UVW1 and UVM2. We computed the optical-ultraviolet spectral energy distributions (SEDs), to get an estimate of the host galaxy fraction at $\nu_*$, which may be significant, and of the AGN luminosity, $L_{UV}$, following the prescriptions by \citet{Vagnetti2013}. We assumed each optical/UV SED in Fig.~\ref{SED} to be the sum of an AGN spectrum, proportional to the average SED by \citet{Richards2006} with a typical slope $\alpha_{AGN}=-0.57$ (with $F_\nu\sim\nu^{\alpha_{AGN}}$), and a host galaxy contribution whose spectrum was modelled with a typical slope $\alpha_{galaxy}=-3$ (see e.g., \citealt{Lusso2010,Vagnetti2013}).
The host galaxy fraction $f_g$ at $\nu_*$ was then derived at each epoch as a function of the sole spectral index. Indeed, expressing the total luminosity as $L=L_*[f_g(\nu/\nu_*)^{-3}+(1-f_g)(\nu/\nu_*)^{-0.57}]$, the spectral index in $\nu_*$ is then equal to $\alpha=[d\log L/d\log\nu]_{\nu=\nu_*}=-3f_g-0.57(1-f_g)$. The fraction $f_g$ was thus estimated inverting the previous relation as $f_g=-(0.57+\alpha)/2.43$. Therefore, when $f_g$ tends to zero then the slope $\alpha$ tends to -0.57, as per the \cite{Richards2006} AGN SED, while when $f_g = 1$ $\alpha$ tends to -3, consistent with a pure host galaxy spectrum.\\
\indent The monochromatic fluxes for the optical and ultraviolet filters were computed by converting the observed rates using the appropriate conversion factors. The spectral index was derived by least squares on these data. Figure~\ref{SED} shows the optical-ultraviolet SEDs for the {\it XMM-Newton} and {\it Swift} observations. From 2007 to recent pointings, the ESO511-G030 SEDs underwent dramatic spectral changes. The optical/UV ESO 511-G030 SEDs appear quite steep with the exception of the 2007 exposure and those from {\it Swift} in 2022. The corresponding spectral index lies in the range between -2.7 and -3. These values are far steeper than the more typical slope of -0.57 derived from the average spectral energy distribution of a statistically significant number of AGNs by Richards et al. (2006) around $\nu_*$.
The variation in the SEDs suggests that the nuclear emission changes, while a substantial constant contribution from the host galaxy is also present. In particular, the steep slopes derived for the observations taken after 2019 but before 2022 can be ascribed to the dominant shape of the host galaxy.
The host galaxy luminosity at 2500\AA\ is then defined as the average value of its estimates at the different epochs, which is found to be $\log L_{galaxy} = 27.68$ (erg s$^{-1}$ Hz$^{-1}$), with a small dispersion, $\sigma = 0.05$. This average value was then subtracted from the total luminosity at 2500\AA\ to obtain the AGN luminosity $L_{UV}$ at each epoch. This luminosity $L_{UV}$ was often smaller than the host galaxy luminosity, as shown in the top panel of Fig.~\ref{aox}. For most observations, the ratio between the AGN and the total monochromatic luminosity (AGN+host) is in the range 0-50\%. In many cases, the SED slope is $\lesssim -3$, the AGN fraction is negligible, and thus not plotted in Figs~\ref{aox} and~\ref{aox1}. On the opposite, the SED of the 2007 archival observation is flatter than the one by \citet{Richards2006} with $\alpha = -0.44$, thus, for this observation, we assumed the total monochromatic luminosity at 2500\AA\ to be due to the AGN only.\\
\indent The $\alpha_{OX}$ derived for our source can be also compared with the well-known $L_{UV}-\alpha_{OX}$ anti-correlation \citep[][]{Vignali2003,Just2007,Vagnetti2010}. Fig.~\ref{aox1} shows the track of ESO 511-G030 in the $\log L_{UV}-\alpha_{OX}$ plane.

\section{Ultraviolet to X-ray modelling}

\begin{figure}[ht]
	\centering
	\includegraphics[width=0.49\textwidth]{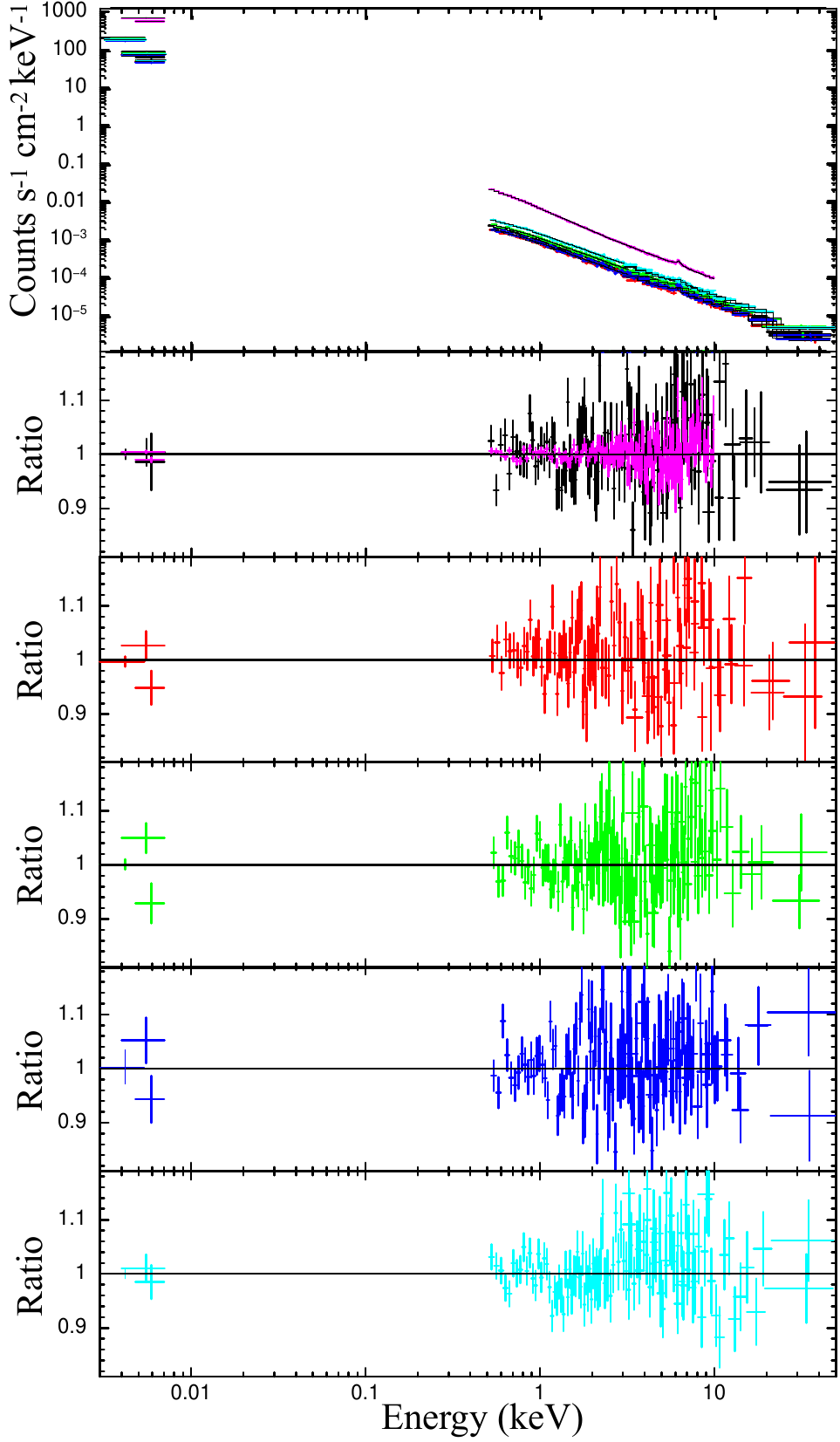}
	\caption{\small{\textit{AGNSED} reproducing the 2019 campaign and the archival \textit{XMM-Newton} exposure (magenta). In bottom panels are shown data-model ratios for each datasets.}}\label{agnsedfit}
\end{figure}

In accordance with previous Sect. 6, the multi wavelength properties of ESO 511-G030 varied dramatically on time. In 2007, in fact, the source ultraviolet-to-X-rays was consistent with the one of a bare Type 1 AGN. Later, once re-observed in 2019, neither the UVs nor the X-rays where compatible with their historical fluxes. Moreover, the ultraviolet emission which in 2007 was fully ascribable to the accretion process, turned out to be galaxy dominated, with only a few percent of the flux being due to the AGN.\\
To better understand the interplay among the different emission components in ESO 511-G030 across the years, we modelled the {\it XMM-Newton} spectrum taken in 2007 and those obtained in 2019. At this stage we included the corresponding \textit{OM} data and tested \textit{AGNSED}, \citep[][]{Kubota2018}. This models allows us to self-consistently reproduce the UV-to-X-ray spectra of ESO 511-G030. In accordance with \cite{Done2012}, \textit{AGNSED} accounts for three distinct emitting regions: an outer standard disc region; a warm Comptonising corona; and the inner hot Comptonising plasma. The flow is radially stratified and emits as a standard disc black body from R$_{\rm out}$ to R$_{\rm warm}$, as warm Comptonisation from R$_{\rm warm}$ to R$_{\rm hot}$ \citep[adopting the passive disc scenario by][]{Petrucci2018} and, below R$_{\rm hot}$ down to R$_{\rm ISCO}$ as the typical hot Comptonisation component.
We thus tested the model\footnote{We did not include in the fit the Balmer continuum nor the Fe II emission lines. In 2019, in fact, the UV emission of ESO511-G030 can be likely ascribed only to the host.}:
\begin{equation}
\rm tbabs_G \times redden\times(galaxy+AGNSED+relxillcp+Borus).\\
\end{equation}
The tbabs$_{\rm G}$ was fixed to the value N$_{\rm H}$=4.33$\times10^{20}$ cm$^{-2}$ \citep{HI4PI} while the \textit{redden} component fixed to a value of E(B-V)=0.056, in agreement with \citet{Schlafly2011}. We tested this model by fitting each observation separately and, within \textit{AGNSED} we allowed the photon indices of the warm and hot coronae to vary as well as R$_{\rm hot}$ and R$_{\rm warm}$. The hot coronal temperature was fixed to 100 keV to fit the 2007 data because there is no constraint above 10 keV, while it was left free to vary in the 2019 data. The warm coronal temperature was computed in both the datasets. We used a co-moving distance of 96 Mpc (derived from the redshift\footnote{\url{https://www.astro.ucla.edu/~wright/CosmoCalc.html}} and adopted the mass for the SMBH in ESO 511-G030 by \citet{Ponti12}. \textit{Relxillcp} and \textit{Borus} (\textit{borus11\_v190815a.fits}), both accounting for a Comptonised continuum, were set similarly to what described in previous Sect. 4.2. These different flavours of \textit{relxill} and \textit{Borus} were used in order to tie the model parameters accounting for a Comptonised continuum (e.g. kT$_{\rm e}$ instead of E$_{\rm cut}$) with the corresponding parameters in \textit{AGNSED}, which assumes an underlying Comptonised continuum and not simply a cut-off power law. 
In the fit procedure we assumed the spin of the central SMBH to be maximally rotating in both {\it relxillcp} and \textit{AGNSED} and the same disc inclination of 30\degree~was also assumed. Then within \textit{AGNSED} the upper limit of the scale height for the hot Comptonisation component, HTmax was set to be 10 R$_{\rm g}$, which mimics a spherical Comptonisation region of similar radius. On the other hand, in relxillcp we set the emissivity profile to its default value of 3. The inner radius for the relxillcp reflection is fixed to R$_{\rm hot}$. This assumption is discussed later. Then the outer disc radius was set to be the same (R$_{\rm out}$=400 R$_{\rm g}$) between the model component \textit{AGNSED} and \textit{relxillcp}. After preliminary tests, we fixed the \textit{relxillcp} parameter R$_{\rm in}$ to 100 R$_{\rm g}$ as it could not be constrained in the 2019 data.
Finally, we added a galaxy template accounting for the host galaxy \citep[matching the morphological type of ESO 511-G030, i.e. Sc,][]{Lauberts1982} contributing to the UV flux. The table was computed following \citet{Ezhikode2017} and included within \textit{XSPEC} as a template named \textit{hostpol} \citep[][]{Polletta2007}. We notice that our galaxy model has a spectral shape between 3-10 eV of $\sim\Gamma$=4. This shape is fully consistent with our assumption in previous Sect. 7 for the host galaxy. We tied the normalisation of \textit{hostpol} across the five spectra.\\
\indent This procedure led us to the best-fit and model to data ratios shown in Fig.~\ref{agnsedfit}. The fit information as well as the corresponding yielded quantities are quoted in Table~\ref{AGNSED}.

\begin{table*}
	\centering
	\setlength{\tabcolsep}{2.5pt}
	\caption{\small{Best-fit parameters derived adopting \textit{AGNSED} on the \textit{XMM-Newton} 2007 exposure and obs. 3 from the 2019 monitoring campaign. The symbol $\star$ is used for those values that were kept fixed in the fitting procedure}\label{AGNSED}}
	\begin{tabular}{c c c c c c c c c c}
		Comp. & Par. & Obs. 1  & Obs. 2 & Obs. 3 & Obs. 4 & Obs. 5 & 2007 & units\\
		\hline
		Galaxy&Norm &4.5$\pm$1.5 & & & & &- &$\times 10^{3}$\\

    \textit{relxillcp}& r$_{\rm in}$ &100$\star$&100$\star$ &100$\star$&100$\star$&100$\star$& 30$\pm$15&R$_{\rm g}$\\
    & $\log \xi$ &1.6$\pm$0.2 & 1.7$\pm$0.2& 1.5$\pm$0.2& 1.4$\pm$0.3&1.6$\pm$0.1 & 1.2$\pm$0.1& \\
    & Norm &<2.0 &1.2$\pm$0.8 &2.1$\pm$0.7 &<1.7 &4.0$\pm$1.0& 25$\pm$1 & $\times 10^{-6}$ ph keV$^{-1}$ cm$^{-2}$ s$^{-1}$\\

    Borus& Norm &1.8$\pm$0.3& 1.9$\pm$0.4&1.6$\pm$0.4 &1.8$\pm$0.4 &1.7$\pm$0.3 & (3.7$\pm$0.1)&$\times 10^{-3}$ ph keV$^{-1}$ cm$^{-2}$ s$^{-1}$\\
    & $\rm \log N_{ H}$&23.2$\pm$0.2&23.2$\pm$0.2&23.4$\pm$0.2&23.2$\pm$0.1&23.2$\pm$0.2&23.7$\pm$0.3&1/(cm$^{-2}$)\\

    \textit{agnsed}&$\log \dot m$ &-2.73$\pm$0.12 &-2.71$\pm$0.11 &-2.66$\pm$0.06 &-2.73$\pm$0.08 &-2.56$\pm$0.02 & -1.65$\pm$0.01&\\
		&$\Gamma_{\rm hot}$ &1.75$\pm$0.02& 1.75$\pm$0.02& 1.76$\pm$0.02&1.75$\pm$2& 1.76$\pm$0.01 &1.91$\pm$0.02 &\\
		&kT$_{\rm hot}$& 60$^{+10}_{-25}$&70$^{+10}_{-40}$&>10 &>30 &>20 & 100$\dagger$ &keV\\
		&R$_{\rm hot}$  & 39$\pm$5 & 38$\pm$6& 42$\pm$3& 40$\pm$3& 41$\pm$3& 27$\pm$1 & R$_{\rm g}$\\
		&$\Gamma_{\rm warm}$ &<2.56 &<2.6 &<2.7 & <2.9&<2.7 &2.64$\pm$0.03 &\\
		&kT$_{\rm warm}$&0.16$\pm$0.06 &0.22$\pm$0.07 &0.20$\pm$0.09 &0.23$\pm$0.07 &<0.3 &0.17$\pm$0.02 &keV\\
		&R$_{\rm warm}$ & <52& <54&<52 &50$\pm$5 &<57 &150$\pm$2 &R$_{\rm g}$\\

        &$\chi^2$/d.of.&280/260&290/256&310/266&320/287&380/331&200/165&\\
        &p.null&0.2&0.1&0.03&0.1&0.05&0.04&\\
	\end{tabular}
\end{table*}

According to these fits, the SED of ESO 511-G030 varied dramatically from 2007 to 2019 as for the Eddington ratio that varied from a value of L/L$_{\rm Edd}\sim$2\% (in the 2007 spectrum soft-excess+power-law), to a rather low and radiatively inefficient value of $\sim$0.2\% in 2019. The standard picture assumed within {\it AGNSED} i.e. the presence of a hot plasma (for R between R$_{\rm isco}$ and R$_{\rm hot}\sim$27 R$_{\rm g}$), a warm one (from R$_{\rm hot}$ to R$_{\rm warm}\sim$150 R$_{\rm g}$) and an accretion disc radially segregated agrees with the 2007 \textit{XMM-Newton} observation.
As said before, in our model the reflection is produced beyond R$_{\rm hot}$. However in \textit{AGNSED} a warm-corona is present between R$_{\rm hot}$ and R$_{\rm warm}$ and the standard disc starts only beyond R$_{\rm warm}$. While we believe that the impact on our best fit results should be limited, our modelling does not take into account the presence of the warm corona in the reflection computation and thus does not provide a fully self-consistent physical picture of the emission emerging from ESO511-G030. To our knowledge the model {\it REXCOR} \citep[][]{Xiang2022} is the sole publicly available model  that self-consistently computes the reflection spectrum  from the combination of an outer standard disk and inner warm corona. However, this model cannot be extended to the UV energy range and cannot be used to fit simultaneously the {\it OM} data.
Finally, the lack of a substantial soft-excess during the 2019 campaign leads to a different picture where the hot corona is now more extended than before and no clear indication of a warm Comptonising region is found. This warm region has shrunk and the disc already extends from $\lesssim$ 50 R$_{\rm g}$, close to the hot component. The relativistic reflection component is less prominent and no constraints on the inner radius were obtained. The cold reflection component is compatible among the 2007 and 2019 data, in agreement with its distant origin from the central engine. Finally, the change for the hot and warm plasma components is also accompanied by the spectra evolving from a softer to a harder state, from $\Gamma$=1.91$\pm$0.03 to and average value of $\Gamma$=1.75$\pm$0.02. 

\section{Discussion and conclusions}

\indent We reported on the spectral and temporal properties of ESO 511-G030 that showed significant variability in both the optical-ultraviolet and X-ray. Our analysis revealed the ESO 511-G030 spectrum to be consistent with a primary power-law $\Gamma$=1.73$\pm$0.02 accompanied by a poorly constrained high energy roll over. The reflected flux we observed in the X-rays of ESO 511-G030 emerges from regions of different densities (Compton-thin and Compton-thick). In Fig.~\ref{borusrelxill}, we find the high energy spectrum to be dominated by reflection off a Compton-thin medium N$_{\rm H}\sim1.8^{23}$ cm$^{-2}$ that does not produce a relevant Compton-hump but accounts for a narrow Fe K$\alpha$ line. The relativistic reflection component reproduces the moderate broad shape of the same emission line and contributes to the overall spectral curvature.\\
\indent Testing our Model C on archival data \citep[our Sect. 5 and the work by][]{Ghosh2020} revealed the primary component to be harder in 2019 than in past observations. These harder states were found in correspondence to lower flux levels, this suggesting the commonly observed softer-when-brighter trend. The reflected flux also varied; in particular the relativistic reflection is found to follow the variations of the primary emission (in agreement with this flux being released in the close surroundings of the central engine), while a less variable behaviour is observed for the cold reflection.\\
\indent One of the main features of the 2019 observational campaign is the lack of a substantial soft-excess. A simple power-law, in fact, dominates the soft-to-hard X-ray spectrum of ESO 511-G030 at least from 2019. The lack of a soft-excess, or its negligible contribution to the overall emission spectrum of ESO 511-G030, is further supported by the analysis of the 2007 and 2019 excess variance spectra. Two different components are in fact needed to account for the 2007 F$_{\rm var}$ spectrum, one responsible for the changes in the X-ray continuum and a second accounting for the soft-excess. On the other hand, the F$_{\rm var}$ computed for the 2019 \textit{XMM-Newton} exposures only requires
a single component accounting for variance due to the nuclear continuum. It is worth noting that, while the F$_{\rm var}$ spectra of Fig.~\ref{fvarfit} sample slightly different timescales (see Sect. 3), we verified that when cutting the 2007 observation into shorter segments (so as to sample similar timescales as for the 2019 F$_{\rm var}$ spectra), a soft excess component still appears in the F$_{\rm var}$ of the lowest flux segment.\\
\indent The absence of a strong soft-excess is quite unusual, since it is ubiquitously observed in AGNs \citep[e.g.][]{Piconcelli2005,Bian09,Gliozzi2020}. Note that ESO 511-G030 data do not require any absorbing component as also discussed in \cite{Laha2014}.\\
\indent The case of this Seyfert galaxy is peculiar and, we can only compare its behaviour with Mrk 1018. This other AGN had been studied in depth by \cite{Noda2018} who observed very different spectral shapes corresponding to different Eddington ratios. From a typical type 1 spectrum with a strong soft-excess, the source dimmed down, became harder and showed a weaker soft-excess. This spectral transition corresponded to a change in the Eddington ratio from L/L$_{\rm Edd} \sim$2\% to L/L$_{\rm Edd} \sim$0.4\% very similar to what is observed here for ESO 511-G030.  As the soft excess is responsible for most of the ionising photons, the dramatic drop in the X-rays also led to the disappearance of the BLR, with this producing the ‘changing-look’ phenomenon. In other words, the presence(lack) of the soft-excess corresponded to a general softening(hardening) of the X-ray continuum emission with an accompanying dramatic change in the disc emission and a disappearance of the optical broad lines.\\
\indent Similar to Mrk 1018, ESO 511-G030 had a dramatic change in its accretion rate passing from L/L$_{\rm Edd} \sim$2\% in 2007 down to L/L$_{\rm Edd} \sim$0.2\% in 2019. This dimming was also accompanied with a dramatic change in the UV SEDs (see Fig.~\ref{SED} and Fig.~\ref{agnsedfit}), though it did not lead to a 'complete' changing look process. An optical Floyds spectrum was, in fact, taken quasi-simultaneously with the \textit{XMM-Newton-NuSTAR} campaign to check whether broad lines were present or not. From a quick comparison between the Floyds spectrum and an 6dF archival one taken in 2000, the H$\beta$ line does not disappear in 2019 with its velocity width being similar between the spectra (FWHM$_{{\rm H}\beta} \sim 4500$ km s$^{-1}$, private communications with Keith Horne and Juan V. Hern\'andez Santisteban)\\
\indent The strong decrease of the accretion rate between 2007 and 2019 seems to be the crucial element to explain the observed spectral UV-X-ray behavior. Indeed this decrease naturally explains the strong decrease of the UV emission, see Fig.s ~\ref{aox} and \ref{aox1}. To 0th order, the decrease of the UV flux would also mean a decrease of the soft photons flux entering and cooling the hot corona. So we would expect an increase of the hot plasma temperature and a hardening of the X-ray spectrum with respect to 2007. The spectral hardening is observed and the absence of a stringent high energy cut-off signature in 2019 agrees with a large corona temperature, much larger than the usual values observed in Seyfert galaxies \citep[e.g.][]{Fabi15,Fabi17,Tamborra2018,Middei2019}. The absence of high signal high energy observations in the archives prevent any comparison with past observations that would help to support this scenario. But the strong decrease of the accretion rate could also explain the absence of the soft X-ray excess in 2019 at least in the case of the warm corona model. The observations agree with the warm corona being the upper layers of the accretion disc \citep[][and references therein]{Petrucci2018}. More importantly, to reproduce the soft X-ray spectral shape, simulations show that a large enough accretion power has to be released inside this warm corona and not in the accretion disc underneath \citep[][]{Rozanska2015,Petrucci2020,Ballantyne2020}. So if the accretion power becomes too low the warm corona cannot be energetically sustained. It is less obvious to understand why the soft X-ray excess would disappear if it is due to relativistically blurred ionised reflection. It is possible however that at low accretion rate the disc becomes more optically thin (or even recedes) producing less reflection as indeed observed in 2019.\\
\indent However, other possible explanations for the lack of the soft-excess in ESO 511-G030 can be viable and new exposures, possibly performed during the awakening of this component, are mandatory in order to shed light onto the engine in this Seyfert galaxy. The increasing UV and X-ray fluxes observed by \textit{Swift} in the first quarter of 2022 encourage us to ask for more observing time to, possibly, observe the revenant soft-excess of ESO 511-G030.

\begin{acknowledgements}

RM thanks Francesco Saturni, Mauro Dadina and Emanuele Nardini for useful discussions and insights. Fondazione Angelo Della Riccia for financial support and Universit\'e Grenoble Alpes and the high energy SHERPAS group for welcoming him at IPAG. Part of this work is based on archival data, software or online services provided by the Space Science Data Center - ASI.  This work has been partially supported by the ASI-INAF program I/004/11/5. RM acknowledges financial contribution from the agreement ASI-INAF n.2017-14-H.0. SB and EP acknowledge financial support from ASI under grants ASI-INAF I/037/12/0 and n. 2017-14-H.O. POP acknowledges financial support from the CNES, the French spatial agency, and from the PNHE, High energy programme of CNRS. ADR  acknowledges financial contribution from the agreement ASI-INAF n.2017-14-H.O. BDM acknowledges support from a Ram\'on y Cajal Fellowship (RYC2018-025950-I) and the Spanish MINECO grant PID2020-117252GB-I00. This  work  is  based  on  observations obtained with: the {\it NuSTAR} mission,  a  project  led  by  the  California  Institute  of  Technology,  managed  by  the  Jet  Propulsion  Laboratory  and  funded  by  NASA; XMM-Newton,  an  ESA  science  mission  with  instruments  and  contributions  directly funded  by  ESA  Member  States  and  the  USA  (NASA).
\end{acknowledgements}
	
\thispagestyle{empty}
\bibliographystyle{aa}
\bibliography{Accepted.bib}

\begin{appendix}
	\section{The Swift-XRT observations}
Results of the spectral analysis performed on the {\it Swift-XRT} observations belonging to our monitoring campaigns. Details are provided in Sect. 6.

	\begin{table}
		\centering
		\caption{\small{Best-fitted parameters as inferred from a spectral analysis of {\it Swift-XRT} spectra. Soft (0.3-2 keV) and hard (2-10 keV) fluxes are given in units of $\times10^{-12} \rm erg~cm^{-2}~s^{-1}$.}}\label{tableswift}
		\begin{tabular}{c c c c c c}
			Time    &   ObsID   & $\Gamma$  & F$_{\rm Soft}$& F$_{\rm Hard}$  &   Cstat/d.o.f.  \\
			\hline
			2018-12-29&03105115001&2.01$\pm$0.26&1.45$\pm$0.14&1.33$\pm$0.47&20/21\\
			2019-07-20&00088915001&1.54$\pm$0.16&2.05$\pm$1.39&4.41$\pm$2.12&56/57\\
			2019-08-02&00088915002&1.74$\pm$0.16&1.85$\pm$0.54&2.77$\pm$1.13&47/54\\
			2019-08-28&00088915003&1.62$\pm$0.23&1.32$\pm$0.68&2.39$\pm$1.32&35/39\\
			2019-12-20&00088915004&1.63$\pm$0.35&1.37$\pm$0.64&2.29$\pm$1.9&14/16\\
			2019-12-23&00088915005&1.62$\pm$0.08&2.21$\pm$1.12&4.24$\pm$1.47&162/159\\
			2019-12-25&00088915006&1.80$\pm$0.19&1.99$\pm$0.43&2.72$\pm$1.13&25/34\\
			2020-01-20&00088915007&1.66$\pm$0.11&1.41$\pm$0.62&2.53$\pm$0.95&99/112\\
			2020-01-21&00088915008&1.51$\pm$0.19&1.29$\pm$0.95&2.9$\pm$1.59&54/49\\
			2020-02-20&00088915009&1.53$\pm$0.11&2.03$\pm$1.40&4.53$\pm$1.86&137/119\\
			2020-02-21&00088915010&1.66$\pm$0.10&1.94$\pm$0.84&3.46$\pm$1.22&90/114\\
			2020-02-25&00088915011&1.56$\pm$0.38&<6.72&<11.56&3/8\\
			2020-03-20&00088915012&1.64$\pm$0.07&2.77$\pm$1.30&5.14$\pm$1.66&174/212\\
			2020-04-20&00088915013&1.60$\pm$0.06&4.50$\pm$2.42&9.05$\pm$2.91&277/276\\
			2020-05-20&00088915014&1.71$\pm$0.09&2.00$\pm$0.72&3.28$\pm$1.03&175/164\\
			2020-06-20&00088915015&1.60$\pm$0.09&1.82$\pm$0.98&3.59$\pm$1.28&160/159\\
			2020-06-21&00088915016&1.44$\pm$0.19&1.77$\pm$1.58&4.36$\pm$2.55&50/44\\
			2021-03-01&00088915017&1.69$\pm$0.07&3.39$\pm$1.33&5.83$\pm$1.76&215/223\\
			2021-04-03&00088915018&1.59$\pm$0.12&2.80$\pm$1.57&5.48$\pm$2.36&82/84\\
			2021-04-07&00088915019&1.64$\pm$0.06&3.95$\pm$1.90&7.49$\pm$2.41&201/237\\
			2021-05-03&00088915020&1.63$\pm$0.08&2.09$\pm$1.03&3.94$\pm$1.34&195/189\\
			2021-06-03&00088915021&1.51$\pm$0.10&1.58$\pm$1.13&3.61$\pm$1.50&104/123\\
			2021-06-08&00088915022&1.60$\pm$0.15&1.64$\pm$0.92&3.14$\pm$1.42&95/75\\
			2021-07-03&00088915023&1.60$\pm$0.10&2.20$\pm$1.19&4.39$\pm$1.61&139/143\\
			2021-07-08&00088915024&1.60$\pm$0.23&1.68$\pm$0.93&3.07$\pm$1.85&37/36\\
			2021-07-09&00088915025&1.74$\pm$0.23&2.03$\pm$0.57&2.85$\pm$1.48&34/33\\
			2021-07-13&00088915026&1.78$\pm$0.12&3.35$\pm$0.83&4.75$\pm$1.53&82/91\\
			2021-08-03&00088915027&1.49$\pm$0.18&1.16$\pm$0.88&2.61$\pm$1.41&63/61\\
			2021-08-04&00088915028&1.58$\pm$0.16&1.24$\pm$0.73&2.45$\pm$1.19&44/60\\
			2021-08-06&00088915029&1.62$\pm$0.12&1.45$\pm$0.72&2.71$\pm$1.08&97/109\\
			2022-02-20&00088915032&1.67$\pm$0.06&4.0$\pm$0.2& 8.6$\pm$0.7&171/208\\
			2022-02-23&00088915033&1.55$\pm$0.10&4.1$\pm$0.3& 10.0$\pm$1.0&122/102\\
			2022-02-24&00088915034&1.69$\pm$0.07&3.5$\pm$0.2&7.4$\pm$0.7&166/179\\
			2022-03-23&00088915035&1.61$\pm$0.07&3.3$\pm$0.2&7.6$\pm$0.5&183/215\\
		\end{tabular}
\end{table}

\end{appendix}
\end{document}